\documentclass[a4paper,11pt]{article}
\pdfoutput=1 

\usepackage{jcappub} 
\usepackage{mathabx}
\usepackage{mathtools}
\usepackage{appendix}
\usepackage{slashed}
\usepackage{multirow}
\usepackage{array}
\usepackage{cellspace}
\usepackage{makecell}

\usepackage[T1]{fontenc} 

\newcommand{\pderiv}[3][]{\frac{\partial^{#1} #2}{\partial #3^{#1}}}
\title{\boldmath The dark Stodolsky effect: constraining effective dark matter operators with spin-dependent interactions}

\author[]{Guillaume Rostagni and}
\author[]{Jack D. Shergold}

\affiliation[]{Institute for Particle Physics Phenomenology, Department of Physics,\\Durham University,\\Durham, UK}

\emailAdd{guillaume.rostagni@durham.ac.uk}
\emailAdd{jack.d.shergold@durham.ac.uk}

\abstract{
We present a comprehensive discussion of the Stodolsky effect for dark matter (DM), and discuss two techniques to measure the effect and constrain the DM parameter space. The Stodolsky effect is the spin-dependent shift in the energy of a Standard Model (SM) fermion sitting in a bath of neutrinos. This effect, which scales linearly in the effective coupling, manifests as a small torque on the SM fermion spin and has historically been proposed as a method of detecting the cosmic neutrino background. We generalise this effect to DM, and give expressions for the induced energy shifts for DM candidates from spin-$0$ to spin-$\frac 32$, considering all effective operators up to mass dimension-6. In all cases, the effect scales inversely with the DM mass, but requires an asymmetric background. We show that a torsion balance experiment is sensitive to energy shifts of $\Delta E \gtrsim 10^{-28}\,\mathrm{eV}$, whilst a more intricate setup using a SQUID magnetometer is sensitive to shifts of $\Delta E \gtrsim 10^{-32}\,\mathrm{eV}$. Finally, we compute the energy shifts for a model of scalar DM, and demonstrate that the Stodolsky effect can be used to constrain regions of parameter space that are not presently excluded. 

}

\begin{document}
\maketitle
\flushbottom
\section{Introduction}\label{sec:introduction}
There is now overwhelming evidence for dark matter (DM) on both galactic~\cite{Rubin:1970zza, 1975ApJ...201..327R, Rubin:1980zd, Bosma:1981zz, Clowe_2006} and cosmological~\cite{Davis:1985rj,BOSS:2016wmc} distance scales, which is estimated to constitute $\sim26\%$ of the total energy density of the universe~\cite{Planck:2018vyg}. Despite this, the exact nature of DM remains a mystery, with all evidence for its existence coming from its gravitational interactions with visible matter. Nevertheless, the possibility remains that DM could interact with Standard Model (SM) fields non-gravitationally, which could allow us to better study its nature.

In order for new fields in a SM extension to be considered as DM candidates, they must be cold in the present epoch and capable of reproducing the observed relic density, electrically neutral, and unable to decay into SM particles over cosmological timescales. This leaves an overwhelmingly large number of DM candidate theories, which are tedious to constrain individually. Effective field theories (EFTs) are an incredibly powerful tool to constrain DM in a model-independent way~\cite{Liem:2016xpm, Banerjee:2020jun, Criado:2021trs,Aebischer:2022wnl}; by making use of the symmetries of the interaction Lagrangian, EFTs reduce the landscape of underlying theories to a finite number of permitted operators. These operators are typically classified by the spin of the DM particle, along with their mass dimension and coupling to the SM, which can then be constrained and mapped onto the candidate DM theory on a case-by-case basis.

To that end, several experiments have already been set up or proposed to directly search for DM using a variety of techniques, each of which have sensitivity to different ranges of parameter space: scattering on ultracold nuclei~\cite{CDMS:2004ghv, CDMS-II:2009ktb, SuperCDMS:2014cds}; scattering in Xenon time projection chambers~\cite{LUX:2013afz, XENON:2022ltv, XMASS:2018bid}; axion telescopes~\cite{CAST:2017uph,ADMX:2019uok,ADMX:2021nhd,MADMAX:2019pub,IAXO:2019mpb}; scattering in particle accelerators~\cite{Bauer:2020nld}; atom interferometers~\cite{Badurina:2019hst, AEDGE:2019nxb, Coleman:2018ozp}. At the same time, DM has also been indirectly constrained using a variety of astrophysical~\cite{AMS:2021nhj, Weisskopf:2000tx, Struder:2001bh,MAGIC:2021mog,HESS:2020zwn,Zitzer:2017xlo,LIGOScientific:2021ffg,Clowe:2006eq} and cosmological~\cite{Planck:2018vyg,WMAP:2012nax,Beutler:2011hx,BOSS:2016wmc} probes.

In this paper, we propose two experiments to observe spin-dependent energy shift induced by a DM background, which is more commonly known as the Stodolsky effect, and has historically been discussed in the context of cosmic neutrino background (C$\nu$B) detection~\cite{Stodolsky:1974aq, Duda:2001hd, Domcke:2017aqj, Bauer:2022lri}. The Stodolsky effect has several features which make a promising avenue for DM detection. First, unlike scattering, the magnitude of the energy shift depends on the DM-SM coupling linearly rather than quadratically, leading to an effect which is less suppressed by tiny coupling constants. Second, whilst many detection techniques depend heavily on the mass of the DM particle under consideration, the Stodolsky effect depends primarily on the velocity of the background particle. For neutrinos, this leads to an energy shift that is largely independent of the neutrino mass~\cite{Bauer:2022lri}, which if also true for dark matter would allow us to probe a wide region of parameter space. On the contrary, the Stodolsky effect for neutrinos requires either a neutrino-antineutrino or left-right helicity asymmetry in the background, the former of which is expected to be absent in the standard C$\nu$B scenario. As we will see, analogous requirements persist for DM backgrounds, potentially restricting the range of models that can give rise to the Stodolsky effect. Even so, both chiral and asymmetric~\cite{Petraki:2013wwa} models of DM exist, which alongside models with finite chemical potential generate an asymmetry during DM production. We additionally note that there are several mechanisms (\textit{e.g.} finite chemical potential, DM reflection at the surface of the Earth~\cite{Arvanitaki:2022oby}, gravitational potentials~\cite{Baym:2021ksj}) through which either asymmetry may develop post-production.

The remainder of this paper will be structured as follows. In Section~\ref{sec:stodolsky} we will review the Stodolsky effect for neutrinos and introduce the general formalism that will be used throughout. Following this, in Section~\ref{sec:DMops} we will compute the magnitude of the Stodolsky effect for all effective DM operators ranging from spin-0 to spin-$\tfrac{3}{2}$, up to dimension-6. Finally, we will discuss the experimental signatures of the Stodolsky effect and the feasibility of this technique for DM detection in Section~\ref{sec:feasibility}, before concluding in Section~\ref{sec:conclusions}.
%
%
\section{The Stodolsky effect}\label{sec:stodolsky}
We begin by reviewing the Stodolsky effect for the C$\nu$B, which has been discussed in several previous works~\cite{Stodolsky:1974aq, Duda:2001hd, Domcke:2017aqj, Bauer:2022lri}. This will closely follow the formalism of~\cite{Bauer:2022lri}, with the exception that we will more carefully treat the external states as partially localised wavepackets, rather than eigenstates of definite momentum. Additionally, we will assume that the neutrinos are monochromatic in the C$\nu$B reference frame, which is a good approximation when their momentum distribution is narrow.

Working in the mass basis, the effective low energy Hamiltonian density for neutrino-electron interactions after applying a Fierz transformation is
\begin{equation}
    \mathcal H_\mathrm{int}(x) = \frac{G_F}{\sqrt 2} \sum_{i,j} \bar \nu_i \gamma_\mu (1-\gamma^5) \nu_j \bar e \gamma^\mu (V_{ij} - A_{ij} \gamma^5) e,
\end{equation}
where $G_F$ is the Fermi constant, $V_{ij}$ and $A_{ij}$ are the effective vector and axial couplings, respectively, and $i,j \in \{1,2,3\}$ denote the neutrino mass eigenstate. To leading order in $\mathcal{H}_\mathrm{int}$, the energy shift of electron helicity state $h_e$ is given by
\begin{equation}\label{eq:eShiftNu}
    \Delta E_e (\vec{p}_e, h_e)=  \sum_{\nu,i,h_\nu} \sum_{N_\nu}\,\langle e_{h_e}, \nu_{i,h_\nu}|  \int d^3x\,\mathcal{H}_\mathrm{int}(x) | e_{h_e}, \nu_{i,h_\nu} \rangle,
\end{equation}
where $h_e$ and $h_\nu$ denote the electron and neutrino helicities, respectively, whilst $\sum_{\nu}$ is the instruction to sum over neutrinos and antineutrinos. Similarly, $\sum_{N_\nu}$ is a sum over all neutrinos in the background with the degrees of freedom specified by the preceding sum. The external states are incoherent superpositions of momentum eigenstates, defined by~\cite{Ghosh:2022nzo, Blas:2022ovz,Smirnov:2022sfo}
\begin{equation}\label{eq:wavepacket}
    |\psi(p_\psi,x_\psi,h_\psi)\rangle = \int \frac{d^3q_\psi}{(2\pi)^3}\frac{1}{\sqrt{2E_{q_\psi}}}\, \omega_\psi(p_\psi,q_\psi)e^{-i\vec{q}_\psi\cdot \vec{x}_\psi} |\{q_\psi,h_\psi\}\rangle,
\end{equation}
with $\psi \in \{e,\nu\}$, where $E_{q_\psi}$ is the energy of the momentum eigenstate with momentum $q_\psi$ and $\omega_\psi$ is a wavepacket function centred on the momentum $p_\psi$. The wavepacket states are normalised to unity, which also sets the normalisation of $\omega_\psi$. We use relativistic normalisation for the momentum eigenstates
\begin{equation}
    | \{p_\psi,h_\psi\} \rangle = \sqrt{2E_p} \,a^\dagger_\psi(\vec p_\psi, h_\psi) | 0 \rangle,
\end{equation}
where $a^\dagger_\psi(\vec{p}, h)$ is the particle creation operator for species $\psi$ with momentum and helicity
$\vec{p}$ and $h$, respectively, whilst its Hermitian conjugate is the corresponding annihilation operator. We denote the antiparticle creation and annihilation operators with $b^\dagger_\psi$ and $b_\psi$, respectively. These satisfy the anticommutation relations 
\begin{equation}\label{eq:anticommutators}
    \left\{ a_i(\vec p,h), a_j^\dagger( \vec q,h') \right\} = \left\{ b_i(\vec p,h), b_j^\dagger( \vec q,h') \right\} = (2\pi)^3 \delta^{(3)}(\vec p - \vec q) \delta_{ij} \delta_{hh'},
\end{equation}
with all other anticommutators equal to zero. Finally, the fermion field operators are decomposed as
\begin{align}
    \psi(x) &= \int \frac{\mathrm d^3 p}{(2\pi)^3} \frac{1}{\sqrt{2E_p}} \sum_h \left( a_\psi(\vec{p},h) u_\psi(p,h) e^{-ip\cdot x} + b^\dagger_\psi(\vec{p},h) v_\psi(p,h) e^{ip \cdot x} \right), \label{eq:diracF1}\\
    \bar \psi(x) &= \int \frac{\mathrm d^3 p}{(2\pi)^3} \frac{1}{\sqrt{2E_p}} \sum_h \left( a^\dagger_\psi(\vec{p},h) \bar u_\psi(p,h) e^{ip\cdot x} + b_\psi(\vec{p},h) \bar v_\psi(p,h) e^{-ip \cdot x} \right),\label{eq:diracF2}
\end{align}
for Dirac fermions, where $u_\psi$ and $v_\psi$ are positive and negative frequency Dirac spinors, respectively. The corresponding field decompositions for Majorana fermions are found by setting $b_\psi = a_\psi$ in~\eqref{eq:diracF1} and~\eqref{eq:diracF2}. Expanding out~\eqref{eq:eShiftNu}, we find
\begin{equation}
    \begin{split}
    \Delta E_e(\vec{p}_e,h_e) &= \sum_{\nu,i,h_\nu} \sum_{N_\nu} \int d^3x \, d\Pi \, \omega_e(p_e,q_e)\omega^*_e(p_e,q'_e)e^{-i(\vec{q}_e - \vec{q}'_e)\cdot \vec{x}_e} \\
    &\times\omega_{\nu_i}(p_{\nu_i},q_{\nu_i})\omega^*_{\nu_i}(p_{\nu_i},q'_{\nu_i})e^{-i(\vec{q}_{\nu_i} - \vec{q}'_{\nu_i})\cdot \vec{x}_{\nu_i}}\\
    &\times \langle \{q'_e,h_e\},\{q'_{\nu_i},h_{\nu_i}\}| \mathcal{H}_\mathrm{int}(x) |\{q_{\nu_i},h_{\nu_i}\},\{q_e,h_e\}\rangle,
    \end{split}
\end{equation}
where we have introduced the shorthand
\begin{equation}
    d\Pi = \frac{d^3q_e}{(2\pi)^3}\frac{d^3q'_e}{(2\pi)^3}\frac{d^3q_{\nu_i}}{(2\pi)^3}\frac{d^3q'_{\nu_i}}{(2\pi)^3} \frac{1}{\sqrt{2E_{q_e}}}\frac{1}{\sqrt{2E_{q'_e}}}\frac{1}{\sqrt{2E_{q_{\nu_i}}}}\frac{1}{\sqrt{2E_{q'_{\nu_i}}}}.
\end{equation}
In line with~\cite{Ghosh:2022nzo,Smirnov:2022sfo}, we now average $\Delta E_e$ over the regions in which the wavepackets are localised, \textit{i.e.} we take
\begin{equation}
    \Delta E_e(\vec{p}_e,h_e) \to \frac{1}{V^2} \int d^3x_e\,d^3x_{\nu_i}\, \Delta E_e(\vec{p}_e,h_e),
\end{equation}
which allows us to eliminate two of the momentum integrals, giving
\begin{equation}
    \Delta E_e(\vec{p}_e,h_e) = \sum_{\nu,i,h_\nu} \sum_{N_\nu} \frac{1}{4V^2}\int \frac{d^3x\,d^3q_e\,d^3p_{\nu_i}}{(2\pi)^6 E_e E_{\nu_i}}\, |\omega_e(p_e,q_e)|^2 |\omega_{\nu_i}(p_{\nu_i},q_{\nu_i})|^2 \,\langle \mathcal{H}_\mathrm{int}\rangle,
\end{equation}
where
\begin{equation}
    \langle \mathcal{H}_\mathrm{int}\rangle = \langle \{q_e,h_e\},\{q_{\nu_i},h_{\nu_i}\}| \mathcal{H}_\mathrm{int}(x) |\{q_{\nu_i},h_{\nu_i}\},\{q_e,h_e\}\rangle.
\end{equation}
Recalling the normalisation $\int \tfrac{d^3q}{(2\pi)^3}\, |\omega(p,q)|^2 = 1$ allows us to identify $|\omega(p,q)|^2/V$ as the phase space density for a single particle. The sum over all particles in the background can therefore be used to replace the wavepacket functions with momentum distribution functions
\begin{equation}
    \sum_{N_\nu} \frac{|\omega_{\nu_i}(p_{\nu_i},q_{\nu_i})|^2}{V} = n_{\nu}(\nu_{i,h_\nu}) f_{\nu_i}(\vec{q}_{\nu_i}), \quad \frac{|\omega_{e}(p_e,q_e)|^2}{V} = \frac{1}{V}\,(2\pi)^3\delta^{(3)}(\vec{p}_e - \vec{q}_e),
\end{equation}
where $n_\nu(\nu_{i,h_\nu})$ is the number density of background neutrino eigenstate $i$ with helicity $h_\nu$. Finally, after noting that nothing in $\langle \mathcal{H}_\mathrm{int}\rangle$ depends on position and considering an electron at rest in the lab frame, we find
\begin{equation}
    \begin{split}
        \Delta E_e(\vec{0},h_e) &= \frac{1}{4m_e} \sum_{\nu,i,h_\nu} n_{\nu}(\nu_{i,h_\nu})\int\frac{d^3p_{\nu_i}}{(2\pi)^3} f_{\nu_i}(\vec{p}_{\nu_i}) \frac{1}{E_{\nu_i}} \langle \mathcal{H}_\mathrm{int}\rangle\Big|_{|\vec{p}_e| = 0}\\
        &= \frac{1}{4m_e} \sum_{\nu,i,h_\nu} n_{\nu}(\nu_{i,h_\nu}) \left\langle\frac{1}{E_{\nu_i}} \langle\mathcal{H}_\mathrm{int}\rangle \right\rangle,
    \end{split}
\end{equation}
where $m_e$ is the electron mass and the outermost angled brackets denote an averaged quantity, which must be done in order to account for the relative motion of the Earth to the C$\nu$B reference frame. The averaging procedure differs slightly between the C$\nu$B and DM, as we do not know the velocity of the former. We therefore use the flux averages from~\cite{Bauer:2022lri} for the C$\nu$B, whilst those for DM are discussed at length in Appendix~\ref{sec:newAv}.

Expanding out the external states, applying the appropriate anticommutation relations and taking the traces of Dirac spinor chains yields~\cite{Bauer:2022lri}
\begin{equation}\label{eq:hintNu}
    \langle \mathcal{H}_\mathrm{int}\rangle = 2\sqrt{2} G_F A_{ii} m_e h_e \Big[m_{\nu_i} h_{\nu_i} (S_e\cdot S_{\nu_i}) - (S_e\cdot p_{\nu_i})\Big] + f(V_{ii}),
\end{equation}
where $h = \pm 1$ denotes the particle spin eigenvalue, $m_{\nu_i}$ denotes the neutrino mass and $f(V_{ii})$ contains terms that do not depend on the electron spin, which will not contribute to the Stodolsky effect. Note that~\eqref{eq:hintNu} takes the opposite sign for external antineutrino states, whilst for external Majorana neutrino states the expectation value is twice as large. The spin vector for massive fermions is given by
\begin{equation}\label{eq:spin4vec}
    S^{\mu} = \left(\frac{\vec{p}\cdot\vec{s}}{m},\vec{s} + \frac{(\vec{p}\cdot\vec{s})\vec{p}}{m(E+m)}\right),
\end{equation}
for a particle with spin vector $\vec{s}$ in its own reference frame. By inspection, we see that $S$ satisfies $(p\cdot S) = 0$. If we restrict our discussion to helicity eigenstates then $\vec{s}$ will be directed along $\vec{p}$, such that~\eqref{eq:spin4vec} reduces to
\begin{equation}\label{eq:spin4vecReduced}
    S^{\mu} = \left(\frac{|\vec{p}|}{m},\frac{E}{m}\frac{\vec{p}}{|\vec{p}|}\right),
\end{equation}
and we instead identify $h = \pm 1$ with the particle helicity\footnote{For simplicity, we will only consider helicity eigenstates for the remainder of this paper.}. Naturally, we cannot use~\eqref{eq:spin4vecReduced} for a particle at rest.

The energy splitting between the two electron spin states is then found by taking the difference between the energy shifts for each spin state, which after performing the flux averaging on~\eqref{eq:hintNu} gives
\begin{equation}\label{eq:neutrinoSplitDirac}
    \begin{split}
    \Delta E_e^D &= \frac{\sqrt{2}G_F}{3}|\vec{\beta}_\Earth| \sum_i A_{ii} \Big[2 \sum_{s_\nu} (2-|\vec{\beta}_{\nu_i}|^2) (n_\nu(\nu_{i,s_{\nu}}^D) - n_\nu(\bar{\nu}_{i,s_{\nu}}^D) \\
    &+\frac{1}{|\vec{\beta}_{\nu_i}|}(3-|\vec{\beta}_{\nu_i}|^2) (n_\nu(\nu_{i,L}^D)-n_\nu(\nu_{i,R}^D) + n_\nu(\bar\nu_{i,R}^D)-n_\nu(\bar\nu_{i,L}^D))\Big],
    \end{split}
\end{equation}
for a Dirac neutrino background, where the subscripts $L$ and $R$ denote left and right helicity neutrinos, respectively, with $R/L$ corresponding to $h_{\nu_i} = \pm 1 \,(\mp 1)$ for (anti)neutrinos. Additionally, $\vec{\beta}_\Earth$ is the relative velocity between the Earth and C$\nu$B reference frame, which may be time dependent, and $\vec{\beta}_{\nu_i}$ is the lab frame neutrino velocity. For completeness, we note that whilst the term scaling as $|\vec{\beta}_{\nu_i}|^{-1}$ appears divergent, it in fact tends to zero as $|\vec{\beta}_{\nu_i}|\to 0$ as a consequence of a vanishing helicity asymmetry for slow neutrinos\footnote{The apparent divergence is an artefact of the frame transformation, and is discussed at length in Section 5.2 and Appendix B of~\cite{Bauer:2022lri}.}. Similarly, we find for a Majorana neutrino background
\begin{equation}\label{eq:neutrinoSplitMajorana}
    \begin{split}
    \Delta E_e^M = \frac{2\sqrt{2}G_F}{3}|\vec{\beta}_\Earth| \sum_i \frac{A_{ii}}{|\vec{\beta}_{\nu_i}|}(3-|\vec{\beta}_{\nu_i}|^2) (n_\nu(\nu_{i,L}^M) - n_\nu(\nu_{i,R}^M)).
    \end{split}
\end{equation}
We immediately see that the Stodolsky effect for neutrinos requires either a non-zero neutrino-antineutrino or helicity asymmetry, but depends only on the neutrino velocity and scales linearly with $G_F$. These features allow an experiment utilising the Stodolsky effect to probe a vast region of DM parameter space, as the effect is less suppressed than scattering in weakly coupled regions, whilst only depending on the dark matter velocity, $|\vec{\beta}_\mathrm{DM}| \simeq 1.2\times 10^{-3}$~\cite{MarrodanUndagoitia:2015veg}, independent of the DM mass.

We are now ready to move onto the Stodolsky effect for DM, which we will henceforth refer to as the dark Stodolsky effect (DSE) to distinguish it from the effect for neutrinos. By analogy with~\eqref{eq:eShiftNu}, the energy shift of an at rest SM fermion $\psi$ in a DM background will be given by
\begin{equation}\label{eq:mastereq}
    \Delta E_\psi (\vec{0}, h_\psi)= \frac{1}{4 m_\psi}\sum_{\mathrm{d.o.f.}} n_{\mathrm{DM}}\left\langle\frac{1}{E_\mathrm{DM}} \langle \mathcal{H}_\mathrm{int} \rangle\right\rangle,
\end{equation}
where the sum runs over the DM degrees of freedom. For the remainder of this paper we will focus on the object appearing inside the angled brackets, which will typically be some kinematic structure depending on the effective DM operator under consideration. When evaluating these expectation values we will only keep the terms that depend on $S_\psi$, as no other terms will contribute to the DSE. The energy splitting of the two SM fermion spin states can then found by starting with our master equation~\eqref{eq:mastereq}, and then taking the difference in the energy shifts for the two spin states. This will typically enter as an overall factor of two. 
%
%
\section{Effective dark matter operators}\label{sec:DMops}
We now turn our attention to the rich landscape of effective DM operators that can give rise to the DSE. For the remainder of this work, we will consider an effective DM Lagrangian of the form
\begin{equation}
    \mathcal{L}_\mathrm{DM} = \mathcal{L}_\mathrm{SM} + \mathcal{L}_\mathrm{kin} + \mathcal{L}_\mathrm{int},
\end{equation}
where $\mathcal{L}_\mathrm{SM}$ is the complete SM Lagrangian, $\mathcal{L}_\mathrm{kin}$ contains the kinetic and mass terms for the DM field, and $\mathcal{L}_\mathrm{int}$ contains effective SM-DM interaction operators. This will take the form
\begin{equation}
    \mathcal{L}_\mathrm{int} = -\frac{g_{\psi\chi}}{\Lambda^{d-4}} \mathcal{O}_\mathrm{DM}^{\mu\nu\dots} \mathcal{O}_{\mu\nu\dots}^{\mathrm{SM}},
\end{equation}
where $g_{\psi\chi}$ denotes the coupling between the SM fermion and DM field, $\Lambda$ is the new physics scale, and $d$ is the combined mass dimension of the SM and DM effective operators, $\mathcal{O}_\mathrm{SM}$ and $\mathcal{O}_\mathrm{DM}$, respectively. We will only work with Lagrangians that are Lorentz invariant, Hermitian, invariant under the SM gauge group and irreducible by the equations of motion, the procedure for which is discussed in Appendix~\ref{sec:reduction}. By inspection of the expectation value, we immediately see that in order for an operator to contribute to the DSE it must contain at least two copies of the field operator corresponding to each external field. For bosonic DM, this gives a minimum combined mass dimension for $\mathcal{O}_\mathrm{SM}$ and $\mathcal{O}_\mathrm{DM}$ of $d=5$, whilst for fermionic DM, the minimum mass dimension is $d = 6$. As such, we will include all effective DM operators up to $d = 6$. However, we will not consider DM operators with $d > 6$, which become increasingly suppressed by the new physics scale $\Lambda$ with increasing $d$.

For an operator $\mathcal{O}_\mathrm{SM} \sim \bar\psi \Gamma_{\mu\nu\dots} \psi$, after expanding out the field operators and external states, and applying the appropriate (anti)commutation relations, we will find the general form for the expectation value containing a trace over the SM fermion Dirac structure
\begin{equation}
    \langle \mathcal{H}_\mathrm{int} \rangle = \frac{g_{\psi\chi}}{\Lambda^{d-4}} P_\chi^{\mu\nu\dots} \,\mathrm{Tr}[u_\psi \bar{u}_\psi \Gamma_{\mu\nu\dots}],
\end{equation}
where $P_\chi$ contains details of the DM kinematics, which may itself contain Dirac traces, $\Gamma$ denotes some string of gamma matrices, and we have used the shorthand $u_\psi \equiv u_\psi(p_\psi,s_\psi)$. The trace can be simplified in a basis independent way be making use of the identities
\begin{align}
    u(p,h) \bar{u}(p,h') &=  \frac{1}{2}(\slashed{p} + m)(1+h\gamma^5 \slashed{S}) \delta_{hh'}, \\
    v(p,h) \bar{v}(p,h') &=  \frac{1}{2}(\slashed{p} - m)(1+h\gamma^5 \slashed{S}) \delta_{hh'},
\end{align}
with $\slashed{A} \equiv \gamma_\mu A^\mu$ for some general four vector $A$. There are a total of five independent gamma matrix structures that can be included in the fermion trace
\begin{equation}\label{eq:gamStrucs}
    1, \quad \gamma^5, \quad \gamma^\mu, \quad \gamma^\mu \gamma^5, \quad \sigma^{\mu\nu},
\end{equation}
where $\sigma^{\mu\nu} = \tfrac{i}{2}[\gamma^\mu,\gamma^\nu]$, $\gamma^5 = \tfrac{i}{4!}\varepsilon_{\alpha\beta\mu\nu} \gamma^\alpha \gamma^\beta \gamma^\mu \gamma^\nu$ and $\varepsilon_{\alpha\beta\mu\nu}$ is the Levi-Civita symbol. Of these, only some will give rise to expectation values that depend on the SM fermion spin, and the remainder can be neglected. Explicitly, we find
\begin{equation}\label{eq:traces}
    \mathrm{Tr}[u_\psi \bar{u}_\psi \Gamma^{\mu\nu\dots}] = 
        \begin{cases}
            2 m_\psi, \quad & \Gamma = 1, \\
            0, \quad & \Gamma = \gamma^5, \\
            2 p_\psi^\mu, \quad & \Gamma = \gamma^\mu, \\
            2 m_\psi h_\psi S_\psi^\mu, \quad & \Gamma = \gamma^\mu \gamma^5, \\
            -2h_\psi \varepsilon_{\alpha\beta\mu\nu} p_\psi^\alpha  S_\psi^\beta, \quad & \Gamma = \sigma^{\mu\nu}, \\
        \end{cases}
\end{equation}
such that of the five independent gamma matrix structures appearing in~\eqref{eq:gamStrucs}, we only need to consider $\gamma^\mu \gamma^5$ and $\sigma^{\mu\nu}$. There is an additional Lorentz invariant structure that we need to consider,
\begin{equation}
    \varepsilon_{\alpha\beta\mu\nu} P_\chi^{\alpha\beta} \bar{u}_\psi \sigma^{\mu\nu} u_\psi,
\end{equation}
which will clearly depend on $S_\psi$. This can be rewritten in terms of $\gamma^5$ as 
\begin{equation} \label{eq:sigma5}
    -2i P_\chi^{\mu\nu} \bar{u}_\psi \sigma_{\mu\nu}\gamma^5 u_\psi,
\end{equation}
and so we will consider the structure $\bar\psi \sigma^{\mu\nu} \gamma^5 \psi$ as an additional `independent' operator throughout. Finally, we note that there are several operator combinations, \textit{e.g.}
\begin{equation}
    \mathcal{O}_\mathrm{DM} \mathcal{O}_\mathrm{SM} = |\phi|^2 \bar\psi \gamma^5 \psi, 
\end{equation}
containing some complex scalar DM field $\phi$, that couple left to right-chiral SM fermions and appear to be dimension-5. However, in order for the SM component to be gauge invariant under $\mathrm{SU}(2)_L$, we require an additional insertion of the SM fermion mass. As a result, if $\Lambda \gg m_\psi$, the operator will effectively scale as one of dimension-6. 
However, as we do not specify the new physics scale, we will treat such operators as dimension-5 throughout. By extension, we will define the dimension of any operator considered in the remainder of this work as the sum of the mass dimensions of its field content and the number of derivatives.
%
%
\subsection{Spin-0}\label{sec:spin0}
New scalar fields are popular candidates for DM~\cite{Duffy:2009ig,Boehm:2003hm,Boehm:2020wbt}, which typically take the form of axion or Higgs-like particles. Axions are well motivated DM candidates, naturally arising in any extension to the SM where an approximate global symmetry is spontaneously broken, where they play the role of the pseudo Nambu-Goldstone boson associated with the symmetry breaking. On the other hand, Higgs-like extensions to the SM require very few additional parameters. In fact, a real singlet scalar coupled to the SM Higgs is the minimal renormalisable extension to the SM capable of explaining DM~\cite{Burgess:2000yq}. 

In our EFT approach, we will make no reference to the underlying theory and simply consider some complex scalar field $\phi$, for which the corresponding field decompositions are
\begin{align}
    \phi(x) &= \int \frac{\mathrm d^3 p}{(2\pi)^3} \frac{1}{\sqrt{2E_p}} \left( a(\vec{p}) e^{-ip\cdot x} + b^\dagger(\vec{p}) e^{ip \cdot x} \right), \label{eq:scalarF1}\\
    \phi^*(x) &= \int \frac{\mathrm d^3 p}{(2\pi)^3} \frac{1}{\sqrt{2E_p}} \left( a^\dagger(\vec{p}) e^{ip\cdot x} + b(\vec{p}) e^{-ip \cdot x} \right),\label{eq:scalarF2}
\end{align}
and the analogous field decomposition for a real scalar DM candidate is found by setting $b = a$. Unlike neutrinos, the creation and annihilation operators for bosonic DM follow commutation relations
\begin{equation}\label{eq:scalarCommutators}
    \left[ a_i(\vec p), a_j^\dagger( \vec q) \right]= \left[ b_i(\vec p), b_j^\dagger( \vec q) \right] = (2\pi)^3 \delta^{(3)}(\vec p - \vec q) \delta_{ij},
\end{equation}
with all other commutators equal to zero. As it turns out, there is only one scalar operator up to dimension-6 that gives rise to the DSE, with interaction Lagrangian
\begin{equation}\label{eq:scalarEg1}
    \mathcal{L}_\mathrm{int}^{\phi} = -\frac{ig_{\psi\phi}}{\Lambda^2}(\phi^* \overleftrightarrow{\partial_\mu} \phi) (\bar\psi\gamma^\mu \gamma^5 \psi),
\end{equation}
where $\phi^* \overleftrightarrow{\partial_\mu} \phi = \phi^*(\partial_\mu \phi) - (\partial_\mu \phi^*)\phi$. The corresponding Hamiltonian density is found via a Legendre transformation
\begin{equation}\label{eq:hamiltonianScalar}
    \mathcal{H}_\mathrm{int}^{\phi} = \sum_\phi \dot\phi\pderiv{\mathcal{L}_\mathrm{int}^\phi}{\dot{\phi}}  - \mathcal{L}_\mathrm{int}^\phi = \frac{ig_{\psi\phi}}{\Lambda^2}\left(\phi^*(\vec{\nabla} \phi) - (\vec{\nabla} \phi^*)\phi\right) \cdot \left(\bar\psi\,\vec{\gamma}\gamma^5 \psi\right),
\end{equation}
where the sum runs over $\phi$ and $\phi^*$. In a background of pure $\phi$ scalars, the relevant expectation value that contributes to the DSE can be computed using the appropriate field decompositions and commutators to find
\begin{equation}\label{eq:expectationScalar}
    \langle\mathcal{H}_\mathrm{int}^{\phi}\rangle = -\frac{2g_{\psi\phi}}{\Lambda^2} \vec{p}_{\phi} \cdot \left(\bar{u}_\psi\, \vec{\gamma}\gamma^5\, u_\psi\right) = -\frac{4g_{\psi\phi}}{\Lambda^2} m_\psi h_\psi (\vec{p}_\phi \cdot \vec{S}_\psi),
\end{equation}
where in going from the first to the second equality we have used the trace identity given in~\eqref{eq:traces}. If a background of pure $\phi^*$ scalars is considered instead, the expectation value~\eqref{eq:expectationScalar} takes the opposite sign. Plugging this into our master equation~\eqref{eq:mastereq}, we therefore find the energy shift of the SM fermion state with spin $h_\psi$
\begin{equation}\label{eq:shiftPhi1}
    \Delta E_\psi^{\phi} (\vec{0}, h_\psi)= -\frac{g_{\psi\phi}}{\Lambda^2} h_\psi \left\langle\frac{(\vec{p}_\phi \cdot \vec{S}_\psi)}{E_\phi} \right \rangle (n_\phi(\phi) - n_{\phi}(\phi^*)),
\end{equation}
with $n_\phi(\phi)$ and $n_{\phi}(\phi^*)$ the number densities of background species $\phi$ and $\phi^*$, respectively. Replacing the average with the expression given in~\eqref{eq:fav1} yields
\begin{equation}
    \Delta E_\psi^{\phi} (\vec{0}, h_\psi)= -\frac{2g_{\psi\phi}}{\Lambda^2} h_\psi  \beta_\Earth (n_\phi(\phi) - n_{\phi}(\phi^*)),
\end{equation}
where $\beta_\Earth$ is the magnitude of the relative velocity between the laboratory and DM reference frames. By taking the difference between the energy shift for each SM fermion spin state, we find an energy splitting
\begin{equation}\label{eq:splitphi1}
    \begin{split}
    \Delta E_\psi^{\phi} &= \Delta E_\psi^{\phi} (\vec{0}, 1) - \Delta E_\psi^{\phi} (\vec{0}, -1)\\ 
    &= -\frac{4g_{\psi\phi}}{\Lambda^2} \beta_\Earth (n_\phi(\phi) - n_{\phi}(\phi^*)).
    \end{split}
\end{equation}
The energy splitting~\eqref{eq:splitphi1} is therefore independent of the DM kinematics, potentially allowing us to constrain scalar DM with masses ranging over many orders of magnitude. Notably, however, we still require a matter-antimatter asymmetry in order to generate a DSE for scalar DM. The culprit in this case is the derivative appearing between the scalar fields in~\eqref{eq:scalarEg1}, which generates an overall minus sign between the positive and negative frequency field modes. This differs from the neutrino case presented in Section~\ref{sec:stodolsky}, where the asymmetry results from the anticommutation relations for fermionic operators. Finally, for completeness we note that the corresponding energy splittings for a real scalar DM background are found by setting $n_\phi(\phi) = n_{\phi}(\phi^*)$, such that~\eqref{eq:splitphi1} should vanish identically. 
\subsection{Spin-$\frac{1}{2}$}\label{sec:spin12}
We now turn our attention to spin-$\tfrac{1}{2}$ dark matter, popular candidates for which include sterile neutrinos~\cite{Dodelson:1993je,Kusenko:2006rh,Petraki:2007gq}, which may also explain short baseline anomalies~\cite{Abazajian:2012ys}, and neutralinos~\cite{Jungman:1995df,Feng:2010gw}, which naturally arise from supersymmetric models.

As we have already seen for neutrinos, the DSE for spin-$\tfrac{1}{2}$ backgrounds differs considerably to the effect for scalar DM, as it can additionally depend on the helicity composition of the background. Furthermore, as the product of four fermion field operators has mass dimension-6, there can be no derivative couplings for fermions at the order considered here. As such, we only need to consider Lorentz structures containing products of linearly independent gamma matrices~\eqref{eq:gamStrucs} and Levi-Civita symbols, the latter of which can be treated as an additional gamma matrix structure, $\sigma^{\mu\nu} \gamma^5$. In all cases, the absence of derivative couplings, along with the anticommutators for fermionic operators will necessarily lead to energy splittings that require a background asymmetry.
\begin{table}[]
    \centering
    \begin{tabular}{Sc|Sc|Sc|Sc}
        Label & $\mathcal{O}_\mathrm{DM}\mathcal{O}_\mathrm{SM}$ & Background & $\langle \mathcal{H}_\mathrm{int}\rangle$ \\ \hline\hline
        \multirow{2.6}{*}{$\mathcal{O}_{\chi_1}$} & \multirow{2.6}{*}{$(\bar\chi \gamma_{\mu} \chi) (\bar\psi \gamma^{\mu}\gamma^5 \psi)$} & $|\chi\rangle$ & $4 m_\psi (p_\chi \cdot S_\psi)$ \\ \cline{3-4} 
         &  & $|\bar\chi\rangle$ & $-4 m_\psi (p_\chi \cdot S_\psi)$ \\  \hline
        \multirow{2.6}{*}{$\mathcal{O}_{\chi_2}$} & \multirow{2.6}{*}{$(\bar\chi \gamma_{\mu}\gamma^5 \chi) (\bar\psi \gamma^{\mu}\gamma^5 \psi)$} & $|\chi\rangle$ & $4 m_\psi m_\chi h_\chi (S_\chi \cdot S_\psi)$ \\ \cline{3-4} 
         &  & $|\bar\chi\rangle$ & $4 m_\psi m_\chi h_\chi (S_\chi \cdot S_\psi)$ \\ \hline
        \multirow{3.8}{*}{$\mathcal{O}_{\chi_3}$} & \multirow{3.8}{*}{$(\bar\chi \sigma_{\mu\nu} \chi) (\bar\psi \sigma^{\mu\nu} \psi)$} & $|\chi\rangle$ & $\begin{aligned}8 h_\chi \big[(p_\chi \cdot S_\psi)&(S_\chi \cdot p_\psi)\\
        &- (p_\chi \cdot p_\psi)(S_\chi \cdot S_\psi)\big]\end{aligned}$ \\ \cline{3-4} 
         &  & $|\bar\chi\rangle$ & $\begin{aligned}-8 h_\chi \big[(p_\chi \cdot S_\psi)&(S_\chi \cdot p_\psi)\\
        &- (p_\chi \cdot p_\psi)(S_\chi \cdot S_\psi)\big]\end{aligned}$ \\ \hline
        \multirow{2.9}{*}{$\mathcal{O}_{\chi_4}$} & \multirow{2.9}{*}{$i(\bar\chi \sigma_{\mu\nu} \chi) (\bar\psi \sigma^{\mu\nu} \gamma^5 \psi)$} & $|\chi\rangle$ & $-8 h_\chi \varepsilon_{\alpha\beta\mu\nu} p_\chi^\alpha p_\psi^\beta S_\chi^\mu S_\psi^\nu$ \\ \cline{3-4} 
         &  & $|\bar\chi\rangle$ & $8 h_\chi \varepsilon_{\alpha\beta\mu\nu} p_\chi^\alpha p_\psi^\beta S_\chi^\mu S_\psi^\nu$
    \end{tabular}
    \caption{Lorentz invariant, Hermitian, gauge invariant and irreducible spin-$\tfrac{1}{2}$ DM operators contributing to the DSE up to dimension-6, along with their corresponding expectation values in a background of Dirac fermions and antifermions, denoted by $|\chi\rangle$ and $|\bar\chi\rangle$, respectively. We leave the global factors of the coupling, new physics scale and SM fermion spin eigenvalue, $h_\psi$, implicit.}
    \label{tab:fermionOperators}
\end{table}

Considering a spin-$\tfrac{1}{2}$ DM candidate $\chi$, we tabulate all irreducible operators contributing the DSE up to dimension-6, along with their corresponding expectation values in Table~\ref{tab:fermionOperators}. For each, we consider the case where the background consists of Dirac $\chi$ and anti-$\chi$, which we denote by $|\chi\rangle$ and $|\bar\chi\rangle$, respectively. The corresponding expectation values in Majorana $\chi$ backgrounds are found by summing those in $|\chi\rangle$ and $|\bar\chi\rangle$ backgrounds. In addition to the operators shown in Table~\ref{tab:fermionOperators}, we could also have considered the operators $\mathcal O_{\chi_5} = i(\bar \chi \sigma_{\mu\nu} \gamma^5 \chi)(\bar \psi \sigma^{\mu\nu} \psi)$ and $\mathcal O_{\chi_6} = (\bar \chi  \sigma_{\mu\nu} \gamma^5 \chi)(\bar \psi \sigma^{\mu\nu} \gamma^5 \psi)$. However, we show in Appendix~\ref{sec:reduction} that these are exactly equal to $\mathcal O_{\chi_4}$ and $\mathcal O_{\chi_3}$, respectively. 

We have already seen the operators $\mathcal{O}_{\chi_1}$ and $\mathcal{O}_{\chi_2}$ in Section~\ref{sec:stodolsky}, which gave rise to the neutrino-antineutrino and helicity asymmetry terms for neutrinos, respectively. The third operator in Table~\ref{tab:fermionOperators}, $\mathcal{O}_{\chi_3}$,  with interaction Lagrangian
\begin{equation}
    \mathcal{L}_\mathrm{int}^{\chi_3} = - \frac{g_{\psi\chi}}{\Lambda^2}(\bar\chi \sigma_{\mu\nu} \chi) (\bar\psi \sigma^{\mu\nu} \psi), 
\end{equation}
leads to an energy shift
\begin{equation}
    \begin{split}
    \Delta E_\psi^{\chi_3}(\vec{0}, h_\psi) = \frac{2g_{\psi\chi}}{m_\psi\Lambda^2}h_\psi\sum_{h_\chi}h_\chi\Bigg[\left\langle\frac{(p_\chi \cdot S_\psi)(S_\chi \cdot p_\psi)}{E_\chi}\right\rangle &- \left\langle\frac{(p_\chi \cdot p_\psi)(S_\chi \cdot S_\psi)}{E_\chi}\right\rangle\Bigg] \\
    &\times (n_{\chi}(\chi_{h_\chi}) - n_{\chi}(\bar\chi_{h_\chi})),
    \end{split}
\end{equation}
which after replacing the averages with~\eqref{eq:fav4} and~\eqref{eq:fav5} yields
\begin{equation}
    \Delta E_\psi^{\chi_3}(\vec{0}, h_\psi) = \frac{7g_{\psi\chi}}{4\Lambda^2} h_\psi  \left(n_{\chi}(\chi_R) - n_{\chi}(\chi_L) - n_{\chi}(\bar\chi_L) + n_{\chi}(\bar\chi_R)\right) + \mathcal{O}\left(\beta_\Earth^2, 1-\beta_r\right),
\end{equation}
where $\beta_r =\beta_\Earth/\beta_c \simeq 1$ is the ratio of the relative frame velocity and galactic circular velocity, $\beta_c$. The resulting energy splitting between the SM fermion spin states has magnitude
\begin{equation}\label{eq:splitChi3}
    \Delta E_\psi^{\chi_3} = \frac{7g_{\psi\chi}}{2\Lambda^2} h_\psi  \left(n_{\chi}(\chi_R) - n_{\chi}(\chi_L) - n_{\chi}(\bar\chi_L) + n_{\chi}(\bar\chi_R)\right),
\end{equation}
to leading order in small quantities, where the subscripts $L$ and $R$ denote the number densities of left and right helicity DM fermions, which satisfy $n_{\chi}(\chi_R) + n_{\chi}(\chi_L) = n_{\chi}(\chi)$ and $n_{\chi}(\bar\chi_L) + n_{\chi}(\bar\chi_R) = n_{\chi}(\bar\chi)$, respectively. Remarkably, this is energy shift is not suppressed by the velocity scale provided that $\beta_r \simeq 1$. Despite this, energy shifts of the form~\eqref{eq:splitChi3} are exceedingly difficult to generate; whilst the helicity asymmetry requirement of~\eqref{eq:splitChi3} naively appears comparable to that of~\eqref{eq:neutrinoSplitDirac} for neutrinos, this is not the case. The first difference is seen when considering Majorana fermions, for which $n_{\chi}(\chi_R) = n_{\chi}(\bar\chi_L)$ and $n_{\chi}(\chi_L) = n_{\chi}(\bar\chi_R)$. This leads to ~\eqref{eq:splitChi3} vanishing identically, whilst~\eqref{eq:neutrinoSplitDirac} becomes~\eqref{eq:neutrinoSplitMajorana}, which importantly is non-zero. The second difference is more subtle. A chiral theory such as the weak interaction will naturally lead to scenarios in which
\begin{equation}
    n_{\chi}(\chi_R) \simeq n_{\chi}(\bar\chi_L) \neq n_{\chi}(\chi_L) \simeq n_{\chi}(\bar\chi_R),
\end{equation}
in particular when the DM fermion is produced relativistically, such that its helicity and chirality coincide\footnote{As helicity is a good quantum number, it is conserved in time. The helicity profile of the DM background today should therefore be the same as at production in the absence of significant late time interactions. See~\cite{Bauer:2022lri} and~\cite{Long:2014zva} for the argument as applied to the C$\nu$B.}. This helicity profile is sufficient to generate a DSE through the operator $\mathcal{O}_{\chi_2}$, but not $\mathcal{O}_{\chi_3}$, which requires a further fermion-antifermion asymmetry (\textit{e.g.} through a chemical potential) to give a non-zero energy splitting. This significantly restricts the number of models that can generate a DSE through operators of the form $\mathcal{O}_{\chi_3}$. Finally, we note that as discussed in Appendix B of~\cite{Bauer:2022lri}, background helicity asymmetries vanish for very cold DM as a consequence of the relative frame velocity. This is true irrespective of the DM spin, and so should be taken into account whenever an operator requires a non-zero helicity asymmetry to contribute to the DSE.

The final operator appearing in Table~\ref{tab:fermionOperators}, $\mathcal{O}_{\chi_4}$, generates an energy shift scaling with
\begin{equation}
    \Delta E^{\chi_4}_\psi(\vec{0},s_\psi) \sim  \varepsilon_{\alpha\beta\mu\nu} p_\chi^\alpha p_\psi^\beta S_\chi^\mu S_\psi^\nu,
\end{equation}
which considering only helicity eigenstates and making use of the identities given in Appendix~\ref{sec:lcIdentity} vanishes identically. We note, however, that this operator may give rise to a non-zero DSE for an alternative experimental setup where the SM fermion is not at rest in the lab frame. 
%
%
\subsection{Spin-1}
Vector bosons remain popular in many models of DM, with candidates including additional $U(1)$ gauge bosons~\cite{ParticleDataGroup:2020ssz,Langacker:2008yv,Bauer:2018onh,Fabbrichesi:2020wbt,Caputo:2021eaa}, superpartners to neutrinos~\cite{Jungman:1995df,Feng:2010gw}, and Kaluza-Klein states in theories with extra dimensions~\cite{Servant:2002aq, Cheng:2002ej,Feng:2010gw}. It is also entirely possible to generate dark hadronic vector states in non-Abelian extensions to the SM~\cite{Garani:2021zrr}. 

The DSE for vector bosons is similar to that for scalar bosons, and may depend on either the total background DM density or require an asymmetry in the presence of derivative couplings. They differ, however, in the fact that vector bosons carry an additional Lorentz index, which expands the number of contributing operators. Here we consider a massive\footnote{In order to be cold, the dark matter background must be massive.} vector field $X^{\mu}$, with field decomposition
\begin{align}
    X_{\mu}(x) &= \int \frac{\mathrm d^3 p}{(2\pi)^3} \frac{1}{\sqrt{2E_p}} \sum_l \left( a(\vec{p},l) \epsilon_{\mu}(p,l) e^{-ip\cdot x} + b^\dagger(\vec{p},l) \epsilon_{\mu}^*(p,l) e^{ip \cdot x} \right), \label{eq:vectorF1}\\
    X_\mu^*(x) &= \int \frac{\mathrm d^3 p}{(2\pi)^3} \frac{1}{\sqrt{2E_p}} \sum_l \left( a^\dagger(\vec{p},l) \epsilon_{\mu}^*(p,l) e^{ip\cdot x} + b(\vec{p},l) \epsilon_{\mu}(p,l) e^{-ip \cdot x} \right),\label{eq:vectorF2}
\end{align}
where the creation and annihilation operators satisfy the commutation relations~\eqref{eq:scalarCommutators}, whilst $\epsilon^\mu(p,l) \equiv \epsilon^\mu_l$ is the polarisation vector with polarisation $l \in \{-1,0,1\}$. Considering the helicity eigenstates for a state with momentum along the $+z$ direction, these take the form
\begin{equation}\label{eq:polVecs}
    \epsilon^\mu(p,1) \equiv \epsilon^\mu_{+} = \frac{1}{\sqrt{2}}{\left(\begin{array}{c}
        0 \\
        1 \\
        i \\
        0
    \end{array}\right)}, \qquad \epsilon^\mu(p,-1) \equiv \epsilon^\mu_{-} = \epsilon^{*\mu}_+, \qquad
    \epsilon^\mu(p,0) \equiv \epsilon^\mu_L = \frac{1}{m}{\left(\begin{array}{c}
        |\vec{p}| \\
        0 \\
        0 \\
        E
    \end{array}\right)},
\end{equation}
which we will refer to as the right, left and longitudinal polarisation states, respectively, together satisfying $\epsilon(p,l) \cdot \epsilon(p,l')^* = -\delta_{ll'}$. The polarisation vectors for momenta along other directions are found by applying the appropriate rotation matrix. These will need to be considered in order to perform the averaging appropriately.   

We tabulate all irreducible operators for vector DM contributing to the DSE up to dimension-6 in Table~\ref{tab:vectorOperators}. As before, we consider each of the cases where the background consists of a complex vector field, $X_\mu$ and its conjugate, $X^*_\mu$, which we denote by $|X\rangle$ and $|X^*\rangle$, respectively. The corresponding expectation values in real $X$ backgrounds are found by summing those in $|X\rangle$ and $|X^*\rangle$ backgrounds.
\begin{table}[]
    \centering
    \begin{tabular}{Sc|Sc|Sc|Sc}
        Label & $\mathcal{O}_\mathrm{DM}\mathcal{O}_\mathrm{SM}$ & Background & $\langle \mathcal{H}_\mathrm{int}\rangle$ \\ \hline\hline
        \multirow{2.6}{*}{$\mathcal{O}_{X_1}$} & \multirow{2.6}{*}{$i (X_\alpha^* \overleftrightarrow{\partial_\mu} X^\alpha) (\bar\psi \gamma^{\mu}\gamma^5 \psi)$} & $|X\rangle$ & $4 m_\psi (\vec p_X \cdot \vec S_\psi)$ \\ \cline{3-4} 
         &  & $|X^*\rangle$ & $-4 m_\psi (\vec p_X \cdot \vec S_\psi)$ \\\hline
        \multirow{3.0}{*}{$\mathcal{O}_{X_2}$} & \multirow{3.0}{*}{$iX_{\mu}^* X_{\nu}(\bar\psi \sigma^{\mu\nu} \psi)$} & $|X\rangle$ & 
        $2\, \mathrm{Im}\left[ \varepsilon_{\alpha\beta\mu\nu} \epsilon^{*\alpha}_{l_X} \epsilon^\beta_{l_X} p_\psi^\mu S_\psi^\nu \right]$ \\ \cline{3-4} 
         &  & $|X^*\rangle$ & 
        $-2\, \mathrm{Im}\left[ \varepsilon_{\alpha\beta\mu\nu} \epsilon^{*\alpha}_{l_X} \epsilon^\beta_{l_X} p_\psi^\mu S_\psi^\nu \right] $ \\ \hline
        \multirow{3.0}{*}{$\mathcal{O}_{X_3}$} & \multirow{3.0}{*}{$X_{\mu}^* X_{\nu}(\bar\psi \sigma^{\mu\nu} \gamma^5 \psi)$} & $|X\rangle$ & 
        $4\, \mathrm{Im}\left[(\epsilon^*_{l_X}\cdot S_\psi)(\epsilon_{l_X}\cdot p_\psi)\right]$ \\ \cline{3-4} 
         &  & $|X^*\rangle$ & 
        $-4\, \mathrm{Im}\left[(\epsilon^*_{l_X}\cdot S_\psi)(\epsilon_{l_X}\cdot p_\psi)\right]$ \\ \hline
         \multirow{3.0}{*}{$\mathcal{O}_{X_4}$} & \multirow{3.2}{*}{$\begin{aligned}i\big[X^{\mu *}(\partial_\mu X_{\nu}) - &(\partial_\mu X_\nu^*)X^\mu\big]\\&\times (\bar\psi \gamma^\nu\gamma^5 \psi)\end{aligned}$} & $|X\rangle$ &
        $-4 m_\psi \mathrm{Re}\left[(\epsilon^*_{l_X}\cdot S_\psi)(\vec{\epsilon}_{l_X}\cdot \vec{p}_X)\right]$ \\ \cline{3-4} 
         &  & $|X^*\rangle$ &
        $4 m_\psi \mathrm{Re}\left[(\epsilon^*_{l_X}\cdot S_\psi)(\vec{\epsilon}_{l_X}\cdot \vec{p}_X)\right]$ \\ \hline
         \multirow{3.0}{*}{$\mathcal{O}_{X_5}$} & \multirow{3.2}{*}{$\begin{aligned}\big[X^{\mu *}(\partial_\mu X_{\nu}) + &(\partial_\mu X_\nu^*)X^\mu\big]\\&\times (\bar\psi \gamma^\nu\gamma^5 \psi)\end{aligned}$} & $|X\rangle$ &
         $4 m_\psi \mathrm{Im}\left[(\epsilon^*_{l_X}\cdot S_\psi)(\vec{\epsilon}_{l_X}\cdot \vec{p}_X)\right]$ \\ \cline{3-4} 
         &  & $|X^*\rangle$ &
         $4 m_\psi \mathrm{Im}\left[(\epsilon^*_{l_X}\cdot S_\psi)(\vec{\epsilon}_{l_X}\cdot \vec{p}_X)\right]$ \\ \hline
         \multirow{3.0}{*}{$\mathcal{O}_{X_6}$} & \multirow{3.0}{*}{$i(X^{*\mu} X_\nu + X^\mu X_\nu^*)(\bar\psi \overleftrightarrow{\partial_\mu} \gamma^\nu \gamma^5 \psi)$} & $|X\rangle$ &
         $-8 m_\psi \mathrm{Re}\left[(\epsilon^*_{l_X}\cdot S_\psi)(\vec{\epsilon}_{l_X}\cdot \vec{p}_\psi)\right]$ \\ \cline{3-4} 
         &  & $|X^*\rangle$ &
         $-8 m_\psi \mathrm{Re}\left[(\epsilon^*_{l_X}\cdot S_\psi)(\vec{\epsilon}_{l_X}\cdot \vec{p}_\psi)\right]$ \\ \hline
         \multirow{2.5}{*}{$\mathcal{O}_{X_7}$} & \multirow{2.5}{*}{$\begin{aligned}i \varepsilon^{\alpha\beta\mu\nu} \big[X_\alpha^*(\partial_\beta X_\mu)-&(\partial_\beta X_\mu^*)X_\alpha\big]\\
         &\times(\bar\psi\gamma_\nu\gamma^5\psi)\end{aligned}$} & $|X\rangle$ &
         $0$ \\ \cline{3-4} 
         &  & $|X^*\rangle$ &
         $0$ \\ \hline
         \multirow{3.0}{*}{$\mathcal{O}_{X_8}$} & \multirow{3.2}{*}{$\begin{aligned}\varepsilon^{\alpha\beta\mu\nu} \big[X_\alpha^*(\partial_\beta X_\mu)+&(\partial_\beta X_\mu^*)X_\alpha\big]\\
         &\times(\bar\psi\gamma_\nu\gamma^5\psi)\end{aligned}$} & $|X\rangle$ &
         $4 m_\psi \mathrm{Im} \left[ \varepsilon_{\alpha\beta i \nu} \epsilon_{l_X}^{*\alpha} \epsilon_{l_X}^\beta p_X^i S_\psi^\nu \right]$ \\ \cline{3-4} 
         &  & $|X^*\rangle$ &
         $4 m_\psi \mathrm{Im} \left[ \varepsilon_{\alpha\beta i\nu} \epsilon_{l_X}^{*\alpha} \epsilon_{l_X}^\beta p_X^i S_\psi^\nu \right]$
    \end{tabular}
    \caption{Lorentz invariant, Hermitian, gauge invariant and irreducible spin-1 DM operators contributing to the DSE up to dimension-6, along with their corresponding expectation values in a background of complex vector bosons and the conjugate field, denoted by $|X\rangle$ and $|X^*\rangle$, respectively. We leave the global factors of the coupling, new physics scale and SM fermion spin eigenvalue, $h_\psi$, implicit.}
    \label{tab:vectorOperators}
\end{table}
The first operator in Table~\ref{tab:vectorOperators}, $\mathcal{O}_{X_1}$, is analogous to the one appearing in $\mathcal{L}^{\phi}_\mathrm{int}$. This has already been discussed in detail in~\ref{sec:spin0}; the only difference here is the overall sign of the energy shift, generated by the contraction of two polarisation vectors. As a result, the energy splitting between the two SM fermion spin states will be the same for $\mathcal{O}_{X_1}$ as for its scalar counterpart, and sensitivity to the individual energy shifts is required to distinguish between the two operators. 

The second operator in Table~\ref{tab:vectorOperators}, $\mathcal{O}_{X_2}$, generates an energy shift
\begin{equation}
    \Delta E_\psi^{X_2}(\vec{0},h_\psi) = \frac{g_{\psi X}}{2m_\psi \Lambda} h_\psi \sum_{l_X} \left\langle\frac{1}{E_X} \mathrm{Im}\left[ \varepsilon_{\alpha\beta\mu\nu} p_\psi^\alpha S_\psi^\beta \epsilon^{*\mu}_{l_X} \epsilon^\nu_{l_X}\right] \right\rangle\left(n_{X}(X_{l_X}) - n_{X}(X^*_{l_X})\right),
\end{equation}
for charged vector bosons, and zero otherwise. We immediately see that the longitudinal modes of $X_\mu$ with real polarisation vectors do not contribute to the DSE. This leaves the remaining two polarisation states, which after substituting in the average~\eqref{eq:favEpsPol} and taking the difference between the two energy shifts gives an SM fermion energy splitting
\begin{equation}\label{eq:splitx2}
    \Delta E_\psi^{X_2} = \frac{7g_{\psi X}}{8 m_X\Lambda}  \left(n_{X}(X_-) - n_{X}(X_+)- n_{X}(X^*_-) + n_{X}(X^*_+)\right),
\end{equation}
to leading order. Similar to~\eqref{eq:splitChi3}, the energy shift from $\mathcal{O}_{X_2}$ is not suppressed by the velocity scale, but requires both a polarisation and matter-antimatter symmetry in order to give a non-zero contribution to the DSE. The former requirement is more difficult for vector bosons than fermions, which permit chiral Lagrangians that preferentially produce fermions of a single helicity at high energies. For vector DM, a polarisation asymmetry must therefore be generated through another mechanism such as scattering on a polarised fermionic background. 
The third operator in Table~\ref{tab:vectorOperators} gives rise to an energy shift
\begin{equation}
    \Delta E_\psi^{X_3}(\vec{0},s_\psi) \sim \left\langle\frac{1}{E_X} \mathrm{Im}\left[(p_\psi \cdot \epsilon_{l_X})(S_\psi \cdot \epsilon^*_{l_X})\right] \right\rangle,
\end{equation}
which vanishes for an SM fermion at rest in the lab frame. We note, however, that there may be a contribution to the energy shift from the right and left polarisation states for other experimental setups. On the other hand, the longitudinal state cannot contribute for any setup as its polarisation vector is real. 

The fourth operator in Table~\ref{tab:vectorOperators} is unique, and leads to an energy splitting
\begin{equation}\label{eq:deX4}
    \begin{split}
    \Delta E_\psi^{X_4} &= \frac{2g_{\psi X}}{\Lambda^2} \sum_{l_X} \left\langle\frac{1}{E_X} \mathrm{Re}\left[ (\epsilon^*_{l_X}\cdot S_\psi)(\vec{\epsilon}_{l_X}\cdot \vec{p}_X)\right] \right\rangle\left(n_{X}(X_{l_X}) - n_{X}(X^*_{l_X})\right) \\
    &= -\frac{4g_{\psi X}}{\Lambda^2}\beta_\Earth \left(n_{X}(X_{L}) - n_{X}(X^*_{L})\right),
    \end{split}
\end{equation}
which depends solely on the density of longitudinally polarised background states. This energy splitting is most closely related to the one generated by $\mathcal{O}_{X_1}$, which instead depends on the total asymmetry between $X_\mu$ and its conjugate. As such, it must always be the case that $\Delta E_{\psi}^{X_1} \geq \Delta E_{\psi}^{X_4}$, which may serve to distinguish the two.

Of the remaining operators, all have vanishing contributions to the Stodolsky effect for our experimental setup: the contribution from $\mathcal{O}_{X_5}$ is proportional to the imaginary part of the kinematic structure found in~\eqref{eq:deX4}, which is real valued after averaging over background momenta; the contributions from $\mathcal{O}_{X_6}$ is proportional to $\vec{p}_\psi$, which is zero for our setup; the contribution due to $\mathcal{O}_{X_7}$ vanishes at the kinematic level, as
\begin{equation}
    \left\langle\mathcal{H}_\mathrm{int}^{X_7} \right\rangle \sim \varepsilon_{\alpha\beta\mu\nu}\,\mathrm{Re}\left[\epsilon^{*\alpha}_{l_X}\epsilon^{\beta}_{l_X}\right] = 0,
\end{equation}
whilst the energy splitting due to $\mathcal{O}_{X_8}$ scales with
\begin{equation}
    \left\langle\mathcal{H}_\mathrm{int}^{X_8} \right\rangle \sim \varepsilon_{\alpha\beta i\nu} \epsilon_{l_X}^{*\alpha} \epsilon_{l_X}^\beta p_X^i S_\psi^\nu,
\end{equation}
which is zero for the longitudinal states since $\epsilon^*_L = \epsilon_L$, and for the right and left helicity states as only their spatial components are non-zero.

Notice that the operator giving rise the Zeeman effect, $\mathcal{O}_F\sim F_{\mu\nu} \bar\psi \sigma^{\mu\nu} \psi$, where $F_{\mu\nu}$ is the field strength tensor, does not appear in Table~\ref{tab:vectorOperators}. This is because it only contains a single copy of the vector field, and as a result has a zero expectation value for incoherent background DM states~\eqref{eq:wavepacket}. Instead, the Zeeman effect occurs in a coherent background, defined as the minimum uncertainty state and by extension the state which is the closest to a classical background. Importantly, bosonic field operators have non-zero expectation values in coherent backgrounds. Such coherent states can be formed by any boson, leading to SM fermion spin-dependent energy shifts that are generated by lower dimension operators than those for incoherent states. It is possible, therefore, that the energy shifts arising from coherent states are significantly larger than those considered here. It is also worth noting that none of the operators in Table~\ref{tab:vectorOperators} describe $U(1)$ gauge bosons, but that the wider class of operators generating energy splittings for coherent backgrounds can. The operator $\mathcal{O}_F$ is such an example. We will explore these states in a future work. 
%
%
\subsection{Spin-$\frac{3}{2}$}
With the exception of the gravitino~\cite{Feng:2010gw}, there are no known spin-$\tfrac{3}{2}$ fermions in renormalisable theories~\cite{Ding:2012sm}. Despite this, spin-$\tfrac{3}{2}$ DM has been shown capable of reproducing the observed relic density~\cite{Garcia:2020hyo}, and can be produced as bound states in non-Abelian extensions to the SM. In particular, the spin-$\tfrac{3}{2}$ baryons are the lightest states of a dark $SU(3)$ with a single quark flavour~\cite{Garani:2021zrr, Antipin:2015xia}. 

Whilst sharing many properties with spin-$\tfrac{1}{2}$ fermions, the additional spin degrees of freedom carried by Rarita-Schwinger (RS) fermions give rise to operators with richer Lorentz structures. This is turn leads a larger number of operators that generate a DSE, the energy shift from which will depend on up to four helicity states. As we will see, the contribution to the energy shift from each helicity state will differ in both sign and magnitude for RS fermions, which may serve as an additional tool to help distinguish them from spin-$\frac{1}{2}$ fermions.

In this section, we consider a spin-$\tfrac{3}{2}$ fermion $\Psi$ with field decomposition
\begin{align}
    \Psi_\mu(x) &= \int \frac{\mathrm d^3 p}{(2\pi)^3} \frac{1}{\sqrt{2E_p}} \sum_{\lambda} \left( a(p,\lambda) \xi_\mu^+(p,\lambda) e^{-ip\cdot x} + b^\dagger(p,\lambda) \xi_\mu^-(p,\lambda) e^{ip \cdot x} \right), \\
    \bar\Psi_\mu(x) &= \int \frac{\mathrm d^3 p}{(2\pi)^3} \frac{1}{\sqrt{2E_p}} \sum_{\lambda} \left( a^\dagger(p,\lambda) \bar{\xi}_\mu^+(p,\lambda)e^{ip\cdot x} + b(p,\lambda) \bar{\xi}_\mu^-(p,\lambda) e^{-ip \cdot x} \right),
\end{align}
where $\lambda \in \{\tfrac{3}{2},\tfrac{1}{2},-\tfrac{1}{2}, -\tfrac{3}{2}\}$ is the helicity of the RS fermion, and again we set $a = b$ for Majorana fermions, whilst
\begin{align}
    \xi_\mu^+(p,\lambda) &= \sum_{\{l,s\}} C_{l,h}^\lambda \epsilon_{\mu}(p,l) u(p,h), \\
    \xi_\mu^-(p,\lambda) &= \sum_{\{l,s\}} C_{l,h}^\lambda \epsilon_\mu^*(p,l) v(p,h),
\end{align}
with the sum running over the values of $l \in \{-1,0,1\}$ and $h = \pm 1$ for which $l + \tfrac{h}{2} = \lambda$. Finally, the Clebsch-Gordan coefficients for an RS field can be found in~\cite{ParticleDataGroup:2020ssz}, and are given by
\begin{alignat}{3}
        \lambda &= +\frac{3}{2}:\quad C^{\frac{3}{2}}_{1,1} = 1, \qquad &\lambda &= +\frac{1}{2}:\quad C^{\frac{1}{2}}_{1,-1} = \sqrt{\frac{1}{3}}, C^{\frac{1}{2}}_{0,1} = \sqrt{\frac{2}{3}}, \\
        \lambda &= -\frac{3}{2}:\quad C^{- \frac{3}{2}}_{-1,-1} = 1, \qquad &\lambda &= -\frac{1}{2}: \quad C^{-\frac{1}{2}}_{-1,1} = \sqrt{\frac{1}{3}}, C^{-\frac{1}{2}}_{0,-1} = \sqrt{\frac{2}{3}},
\end{alignat}
with all other coefficients equal to zero.

Once more, we tabulate all irreducible operators for RS fermion DM contributing to the DSE up to dimension-6 in Table~\ref{tab:rsOperators}. As before, we consider backgrounds of RS fermions, $\Psi$ and anti-RS fermions, $\bar\Psi$, which we denote by $|\Psi\rangle$ and $|\bar\Psi\rangle$, respectively. The corresponding expectation values in backgrounds of RS fermions that satisfy the Majorana condition are found by summing those in $|\Psi\rangle$ and $|\bar\Psi\rangle$ backgrounds. We additionally introduce the shorthand
\begin{equation}
    \sum_C f(l_\Psi,h_\Psi) \equiv \sum_{\{l_\Psi,h_\Psi\}} \left(C^{\lambda_\Psi}_{l_\Psi,h_\Psi}\right)^2 f(l_\Psi,h_\Psi),
\end{equation}
with $f$ some arbitrary function depending on the helicity structure of the background, and note that the argument used to exclude the operators $\mathcal O_{\chi_5}$ and $\mathcal O_{\chi_6}$ in Section~\ref{sec:spin12} applies here to the equivalent operators with spin-$\frac 32$ fields.
\begin{table}[]    \centering
    \begin{tabular}{Sc|Sc|Sc|Sc}
        Label & $\mathcal{O}_\mathrm{DM}\mathcal{O}_\mathrm{SM}$ & Background & $\langle \mathcal{H}_\mathrm{int}\rangle$ \\ \hline\hline
        \multirow{2.6}{*}{$\mathcal{O}_{\Psi_1}$} & \multirow{2.6}{*}{$(\bar{\Psi}_\alpha \gamma_\mu \Psi^\alpha) (\bar\psi \gamma^\mu \gamma^5 \psi)$} & $|\Psi\rangle$ & $-4 m_\psi (p_\Psi \cdot S_\psi)$ \\ \cline{3-4} 
         &  & $|\bar\Psi\rangle$ & $4 m_\psi (p_\Psi \cdot S_\psi)$ \\ \hline
        \multirow{2.6}{*}{$\mathcal{O}_{\Psi_2}$} & \multirow{2.6}{*}{$(\bar{\Psi}_\alpha \gamma_\mu \gamma^5 \Psi^\alpha) (\bar\psi \gamma^\mu \gamma^5 \psi)$} & $|\Psi\rangle$ & $-4 m_\psi m_\Psi \sum_C h_\Psi (S_\Psi \cdot S_\psi)$ \\ \cline{3-4} 
         &  & $|\bar\Psi\rangle$ & $-4 m_\psi m_\Psi  \sum_C h_\Psi (S_\Psi \cdot S_\psi)$ \\ \hline
        \multirow{3.0}{*}{$\mathcal{O}_{\Psi_3}$} & \multirow{3.0}{*}{$i(\bar\Psi_{\mu} \Psi_{\nu})(\bar\psi \sigma^{\mu\nu} \psi)$} & $|\Psi\rangle$ & 
        $4 m_\Psi \sum_C\mathrm{Im}\left[ \varepsilon_{\alpha\beta\mu\nu} p_\psi^\alpha S_\psi^\beta \epsilon^{*\mu}_{l_\Psi} \epsilon^\nu_{l_\Psi}\right] $ \\ \cline{3-4} 
         &  & $|\bar\Psi\rangle$ & 
        $-4 m_\Psi \sum_C\mathrm{Im}\left[ \varepsilon_{\alpha\beta\mu\nu} p_\psi^\alpha S_\psi^\beta \epsilon^{*\mu}_{l_\Psi} \epsilon^\nu_{l_\Psi}\right] $ \\ \hline
        \multirow{4.0}{*}{$\mathcal{O}_{\Psi_4}$} & \multirow{4.0}{*}{$(\bar\Psi_{\alpha} \sigma_{\mu\nu}\Psi^{\alpha})(\bar\psi \sigma^{\mu\nu} \psi)$} & $|\Psi\rangle$ & $\begin{aligned}- 8\, \textstyle{\sum_C} h_\Psi \big[&(p_\Psi \cdot S_\psi)(S_\Psi \cdot p_\psi)\\
         &- (p_\Psi \cdot p_\psi)(S_\Psi \cdot S_\psi)\big]\end{aligned}$ \\ \cline{3-4} 
         &  & $|\bar\Psi\rangle$ & $\begin{aligned}8\, \textstyle{\sum_C} h_\Psi \big[&(p_\Psi \cdot S_\psi)(S_\Psi \cdot p_\psi)\\
         &- (p_\Psi \cdot p_\psi)(S_\Psi \cdot S_\psi)\big]\end{aligned}$ \\ \hline
         \multirow{3.0}{*}{$\mathcal{O}_{\Psi_5}$} & \multirow{3.0}{*}{$(\bar\Psi_{\mu} \Psi_{\nu})(\bar\psi \sigma^{\mu\nu} \gamma^5 \psi)$} & $|\Psi\rangle$ & 
         $8 m_\Psi \sum_C \mathrm{Im}\left[(S_\psi \cdot \epsilon^*_{l_\Psi})(p_\psi \cdot \epsilon_{l_\Psi})\right]$ \\ \cline{3-4} 
         &  & $|\bar\Psi\rangle$ & 
         $-8 m_\Psi \sum_C \mathrm{Im}\left[(S_\psi \cdot \epsilon^*_{l_\Psi})(p_\psi \cdot \epsilon_{l_\Psi})\right]$ \\ \hline
         \multirow{2.8}{*}{$\mathcal{O}_{\Psi_6}$} & \multirow{2.8}{*}{$i(\bar\Psi_{\alpha} \sigma_{\mu\nu}\Psi^{\alpha})(\bar\psi \sigma^{\mu\nu}\gamma^5 \psi)$} & $|\Psi\rangle$ & $8 \sum_C h_\Psi\varepsilon_{\alpha\beta\mu\nu} p_\Psi^\alpha p_\psi^\beta S_\Psi^\mu S_\psi^\nu$ \\ \cline{3-4} 
         &  & $|\bar\Psi\rangle$ & $-8 \sum_C h_\Psi\varepsilon_{\alpha\beta\mu\nu} p_\Psi^\alpha p_\psi^\beta S_\Psi^\mu S_\psi^\nu$ \\
    \end{tabular}
    \caption{Lorentz invariant, Hermitian, gauge invariant and irreducible spin-$\tfrac{3}{2}$ DM operators contributing to the DSE up to dimension-6, along with their corresponding expectation values in a background of RS and anti-RS fermions, denoted by $|\Psi\rangle$ and $|\bar\Psi\rangle$, respectively. We leave the global factors of the coupling, new physics scale and SM fermion spin eigenvalue, $h_\psi$, implicit.}
    \label{tab:rsOperators}
\end{table}

The first operator, $\mathcal{O}_{\Psi_1}$, gives the same energy shift as the similar spin-$\frac{1}{2}$ operator $\mathcal{O}_{\chi_1}$ up to an overall sign, which results from the contraction of two polarisation vectors. It therefore only requires a matter-antimatter asymmetry in order to generate a DSE, but cannot tell us anything about the helicity structure of the background. This naturally makes it difficult to distinguish from $\mathcal{O}_{\chi_1}$. 

The remaining operators are far more interesting. Consider $\mathcal{O}_{\Psi_2}$, which gives rise to an energy shift
\begin{equation}
    \begin{split}
    \Delta E_\psi^{\Psi_2}(\vec{0}, h_\psi) &= -\frac{g_{\psi\Psi}}{\Lambda^2} m_\Psi h_\psi\sum_{\lambda_\Psi} \sum_C h_\Psi\left\langle \frac{1}{E_\Psi} (S_\Psi \cdot S_\psi)\right\rangle\left(n_\Psi(\Psi_{\lambda_\Psi}) + n_\Psi(\bar{\Psi}_{\lambda_\Psi})\right) \\
    &= \frac{7g_{\psi\Psi}}{8\Lambda^2} h_\psi \Big[\left(n_\Psi(\Psi_{++}) + n_\Psi(\bar{\Psi}_{++}) - n_\Psi(\Psi_{--}) - n_\Psi(\bar{\Psi}_{--})\right) \\
    &+ \frac{1}{3}\left(n_\Psi(\Psi_{+-}) + n_\Psi(\bar{\Psi}_{+-}) - n_\Psi(\Psi_{-+}) - n_\Psi(\bar{\Psi}_{-+})\right)\Big],
    \end{split}
\end{equation}
where the subscripts $\pm\pm$ and $\pm\mp$ refer to the $\pm \tfrac{3}{2}$ and $\pm \tfrac{1}{2}$ helicity states, respectively. Taking the difference between the energy shifts for each spin state gives the energy splitting due to $O_{\Psi_2}$
\begin{equation}\label{eq:splitpsi2}
    \begin{split}
    \Delta E_\psi^{\Psi_2} &= \frac{7g_{\psi\Psi}}{4\Lambda^2} \Big[\left(n_\Psi(\Psi_{++}) + n_\Psi(\bar{\Psi}_{++}) - n_\Psi(\Psi_{--}) - n_\Psi(\bar{\Psi}_{--})\right) \\
    &+ \frac{1}{3}\left(n_\Psi(\Psi_{+-}) + n_\Psi(\bar{\Psi}_{+-}) - n_\Psi(\Psi_{-+}) - n_\Psi(\bar{\Psi}_{-+})\right)\Big],
    \end{split}
\end{equation}
which requires a non-zero helicity asymmetry in order to generate a DSE, akin to $\mathcal{O}_{\chi_2}$. This is easily achieved in a chiral theory similar to the weak interaction. Owing to the Clebsch-Gordan coefficients, however, the contribution to the DSE from the $\pm \tfrac{1}{2}$ helicity states is suppressed by a factor of three, which for the same total DM density leads to a reduced energy shift. As such, if the mass, and by extension the number density of the DM is known, the reduced energy splitting could serve as a tool to distinguish between spin-$\tfrac{1}{2}$ and spin-$\tfrac{3}{2}$ DM backgrounds. Although difficult to observe, we also note that the energy shifts of the individual spin states differ by an overall sign between $\mathcal{O}_{\Psi_2}$ and $\mathcal{O}_{\chi_2}$. The operator $\mathcal{O}_{\Psi_3}$ yields a similarly suppressed energy splitting
\begin{equation}\label{eq:splitpsi3}
    \begin{split}
    \Delta E_\psi^{\Psi_3} &= \frac{7g_{\psi\Psi}}{4\Lambda^2}  \Big[\left(n_\Psi(\Psi_{++}) - n_\Psi(\bar{\Psi}_{++}) - n_\Psi(\Psi_{--}) + n_\Psi(\bar{\Psi}_{--})\right) \\
    &+ \frac{1}{3}\left(n_\Psi(\Psi_{+-}) - n_\Psi(\bar{\Psi}_{+-}) - n_\Psi(\Psi_{-+}) + n_\Psi(\bar{\Psi}_{-+})\right)\Big],
    \end{split},
\end{equation}
which is only non-zero in a background with both a fermion-antifermion and helicity asymmetry. In this case, we note the analogous lower spin operator is in fact bosonic, $\mathcal{O}_{X_3}$, which should result in a slightly larger splitting for the same background density. However, the biggest difference is in the generation of~\eqref{eq:splitx2} and~\eqref{eq:splitpsi3}; as previously discussed, a helicity asymmetry cannot arise at the Lagrangian level for bosons, but
are possible in chiral theories of fermions which if relativistic at production prefer a given helicity. Consequently, it is much easier to generate the DSE from $\mathcal{O}_{\Psi_3}$. The remaining three operators, $\mathcal{O}_{\Psi_4}$, $\mathcal{O}_{\Psi_5}$ and $\mathcal{O}_{\Psi_6}$ are analogous to $\mathcal{O}_{\chi_3}$, $\mathcal{O}_{X_3}$ and $\mathcal{O}_{\chi_4}$, respectively, such that only the first contributes a DSE for the experimental setup considered here. In the same way as~\eqref{eq:splitpsi2}, the energy shifts due to $\mathcal{O}_{\Psi_4}$ differ from their analogues by an overall sign and a small suppression factor from the Clebsch-Gordan coefficients. Finally, we have omitted the pseudoscalar analogues of $\mathcal{O}_{\Psi_3}$ and $\mathcal{O}_{\Psi_5}$, proportional to $\Psi_\mu \gamma^5 \Psi_\nu$, from Table~\ref{tab:rsOperators} as the expectation values of their Hamiltonians vanish trivially using~\eqref{eq:traces}.

The discussion here is easily extended to higher spin states, which we naively expect will differ only in the overall sign and magnitude of their DSEs. In particular, the magnitude of the DSE for most operators should decrease with increasing spin, as progressively smaller Clebsch-Gordan coefficients will suppress the contribution from the intermediate helicity states. 
%
%
\section{Experimental feasibility}\label{sec:feasibility}
Observing the tiny energy splittings induced by the DSE directly is a remarkable challenge due to their small magnitude. Take for example the splitting due to the C$\nu$B, whose magnitude is expected to be of order
\begin{equation}\label{eq:splitCNB}
    |\Delta E_\psi| \sim G_F\beta_\Earth n_{\nu,0} \simeq 5\times 10^{-39}\,\mathrm{eV},
\end{equation}
assuming maximal neutrino-antineutrino asymmetry, where we have used $\beta_\Earth \simeq 10^{-3}$ and $n_{\nu,0} = 56\,\mathrm{cm}^{-3}$ is the predicted relic neutrino density per degree of freedom. This is approximately thirty orders of magnitude smaller than the energy splitting due to the Zeeman effect in a $1\,\mathrm{G}$ magnetic field. Clearly then, this effect is nigh impossible to observe on the scale of a single target. To that end, we identify two methods utilising macroscopic targets through which the DSE may be observed. Both of these rely on the same property; as a result of the energy splitting due to the DM background, the SM fermion Hamiltonian, $H_\psi$, and spin operators orthogonal to the DM wind, $S_\perp$, no longer commute, leading to a spin precession
\begin{equation}\label{eq:spinPrecession}
    \frac{dS_\perp}{dt} = i[H_\psi, S_\perp] \sim \mathcal{O}(\Delta E_\psi),
\end{equation}
which can equivalently be interpreted as a torque. A ferromagnet with polarisation transverse to the DM wind will therefore experience a macroscopic acceleration as a result of the spin precession, which can be observed with a Cavendish-style torsion balance. Alternatively, a target initially polarised along an external magnetic field will develop some transverse magnetisation as a consequence of the DM background, which may measurable with a SQUID magnetometer. We will explore each of these methods in turn.
\subsection{Torsion balance}\label{sec:torsion}
The possibility of using a torsion balance to observe the tiny energy splittings due to the C$\nu$B was first identified by Stodolsky in~\cite{Stodolsky:1974aq} and has since been discussed in several works~\cite{Duda:2001hd,Domcke:2017aqj,Bauer:2022lri}. A single SM fermion interacting with the DM background will experience a torque $\tau_\psi \simeq |\Delta E_\psi|$, such that a macroscopic target consisting of $N_\psi$ fermions with degree of polarisation $P$ will experience a total torque
\begin{equation}
    \tau_\mathrm{tot} \simeq P N_\psi |\Delta E_\psi| = \frac{N_A}{m_A}\frac{P}{A} M \times \begin{cases}
        Z |\Delta E_e|, &\quad \psi = e, \\
        |\Delta E_N|, &\quad \psi = N,
    \end{cases}
\end{equation}
where $N$ denotes an atomic nucleus, $N_A$ is the Avogadro number, whilst $M$, $A$ and $Z$ denote the total mass, the mass number and atomic number of the target, respectively. We have additionally introduced the ``Avogadro mass'' $m_A = 1\,\mathrm{g}\,\mathrm{mol}^{-1}$. To estimate the sensitivity of a torsion balance to this energy splitting, we consider the same setup as~\cite{Bauer:2022lri} using a torsion balance consisting of $N_m$ spherical, uniformly dense ferromagnets a distance $R$ away from some central axis. To maximise the sensitivity, we additionally assume that opposing ferromagnets are polarised antiparallel to one another. For this setup, the torsion balance will experience a linear acceleration
\begin{equation}
    a \simeq \frac{N_A}{m_A}\frac{P}{A} \frac{N_m}{R} \times \begin{cases}
            Z |\Delta E_e|, &\quad \psi = e, \\
            |\Delta E_N|, &\quad \psi = N.
        \end{cases}
\end{equation}
As such, if accelerations as small as $a_0$ can be measured, the experiment is sensitive to energy splittings
\begin{equation}
    \begin{split}
    |\Delta E_\psi| &\gtrsim a_0 \frac{m_A}{N_A} \frac{A}{P} \frac{R}{N_m} \times \begin{cases}
        \frac{1}{Z}, &\quad \psi = e, \\
        1, &\quad \psi = N,
        \end{cases}\\
        &=  (5.2\cdot 10^{-28}\,\mathrm{eV}) \left[\frac{a_0}{10^{-15}\,\mathrm{cm}\,\mathrm{s}^{-2}}\right]\left[\frac{R}{1\,\mathrm{cm}}\right]\left[\frac{2}{N_m}\right]\frac{A}{P} \times \begin{cases}
        \frac{1}{Z}, &\quad \psi = e, \\
        1, &\quad \psi = N,
        \end{cases}
    \end{split}
\end{equation}
where for our reference sensitivity we have used $a_0 = 10^{-15}\,\mathrm{cm}\,\mathrm{s}^{-2}$, which has recently been achieved in torsion balance tests of the weak equivalence principle~\cite{Wagner:2012ui}. By comparison with~\eqref{eq:splitCNB}, we see that this torsion balance experiment is insensitive to the C$\nu$B, but may still be able to observe DM for which the background number density $n_\mathrm{DM} \gg n_{\nu,0}$. In particular, as the background DM number density scales as $n_\mathrm{DM} = \rho_\mathrm{DM}/m_\mathrm{DM}$, where $\rho_\mathrm{DM} \simeq 0.4\,\mathrm{GeV}\,\mathrm{cm}^{-3}$ is the local dark matter energy density~\cite{Read:2014qva}, low mass DM scenarios are ideal candidates for detection using this method. Finally, we note that a torsion balance consisting test masses suspended by superconducting magnets has been considered in~\cite{Hagmann:1999kf}, which has an estimated sensitivity to accelerations as small as $a_0\simeq 10^{-23}\,\mathrm{cm}\,\mathrm{s}^{-2}$. This, in turn, would allow us to probe energy splittings of order $10^{-36}\,\mathrm{eV}$. 
\subsection{SQUID magnetometer}\label{sec:squid}
The DM wind resulting from the relative motion of the Earth through the background can acts similarly to a magnetic field, leading to the spin precession~\eqref{eq:spinPrecession}. As such, if the target spins are initially aligned along some fixed external magnetic field $\vec{B}_\mathrm{ext}$ that is not colinear with the DM wind, the presence of the background will cause the spins to shift away from the axis of $\vec{B}_\mathrm{ext}$ and give rise to a small transverse magnetisation. The spins will then precess around the combined magnetic field and DM wind with some characteristic frequency, which can be detected using a highly sensitive SQUID magnetometer. This idea has previously been discussed in the context of axion DM in~\cite{Graham:2013gfa}, and is the basis of the CASPEr experiment~\cite{Budker:2013hfa}.

Following the calculations in Appendix~\ref{sec:squidDeriv}, we find that the transverse magnetisation of a target consisting of $N_\psi$ spins evolves as
\begin{equation}\label{eq:magnetisation}
    |M_\mathrm{\perp}(t)| = \frac{2\rho N_A }{m_A}\frac{P}{A} \frac{|R \sin\left(\frac{\omega_{\psi,0}}{2} t\right)|}{1+R^2} \sqrt{1+R^2\cos^2\left(\frac{\omega_{\psi,0}}{2} t\right)} \times\begin{cases}
        Z \mu_e, &\quad \psi = e, \\
        \mu_N, &\quad \psi = N,
    \end{cases}
\end{equation}
where $\rho$ is the mass density of the target, $\mu_\psi$ denotes the magnetic moment of species $\psi$, $R = \Delta E_\psi/\Delta E_{\psi,B}$ is the ratio of the DM and Zeeman energy splittings, and $\omega_{\psi,0} = \Delta E_{\psi,B} \sqrt{1+R^2}$. In~\eqref{eq:magnetisation} we have assumed that the DM wind is exactly perpendicular to $\vec{B}_\mathrm{ext}$, which maximises the transverse magnetisation, and that both the external magnetic field and DM wind directions are constant in time\footnote{For a given choice of axes, at least one of either the DM wind or magnetic field direction must have some time dependence due to the evolution of the relative velocity between the laboratory and DM reference frames.}. We give the full expression for $|M_\perp(t)|$ and discuss the time dependence in Appendix~\ref{sec:squidDeriv}.   

\begin{figure}
    \centering
    \includegraphics[width = \textwidth]{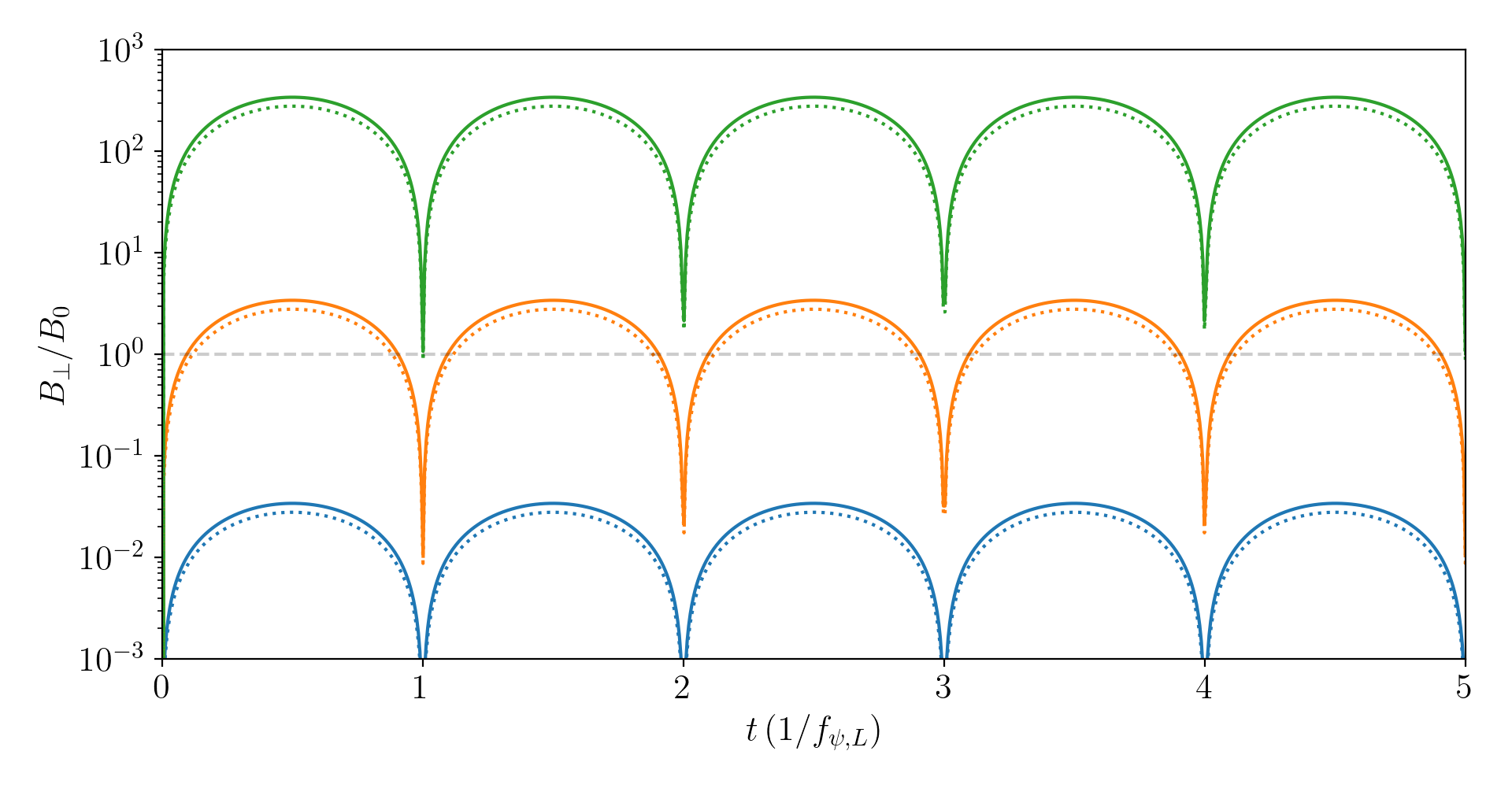}
    \caption{Evolution of the transverse magnetic field generated by the DM background, normalised by the SQUID sensitivity $B_0 = 10^{-16}\,\mathrm{T}$. From bottom to top, the blue, orange and green curves correspond to energy splittings $\Delta {E} = 2\cdot 10^{-32}\,\mathrm{eV}$, $2\cdot 10^{-30}\,\mathrm{eV}$ and $2\cdot 10^{-28}\,\mathrm{eV}$, respectively, whilst the solid and dotted curves correspond in turn to the cases where $\beta_{\Earth,\parallel} = 0$ and $\beta_{\Earth,\parallel} = 1/\sqrt{3}$. Finally, we assume an applied magnetic $|\vec{B}_\mathrm{ext}| = 10^{-10}\,\mathrm{T}$, and an iron target with magnetic moment $\mu_\psi = 3.15\cdot 10^{-8}\,\mathrm{eV}\,\mathrm{T}^{-1}$, equal to the nuclear magneton.}
    \label{fig:squid}
\end{figure}

The transverse magnetisation has a maximum of
\begin{equation}\label{eq:mmax}
    |M_\perp(t_\mathrm{max})| = \frac{2\rho N_A}{m_A}\frac{P}{A}\frac{|R|}{1+R^2} \times \begin{cases}
        Z\mu_e, &\quad \psi = e, \\
        \mu_N, &\quad \psi = N,
    \end{cases} \quad t_\mathrm{max} = \frac{(2k+1)\pi}{\omega_{\psi,0}}, 
\end{equation}
for $R \leq 1$, with $k \in \{0,1,2,\dots\}$. Supposing that a magnetometer can precisely measure transverse magnetic fields with magnitude $B_0$, we will have sensitivity to energy splittings with magnitude
\begin{equation}\label{eq:squidSens}
    \begin{split}
    |\Delta E_\psi| &\gtrsim B_0 |\vec{B}_\mathrm{ext}| \frac{m_A}{\rho N_A} \frac{A}{P} \times \begin{cases}
        \frac{1}{Z}, &\quad \psi = e, \\
        1, &\quad \psi = N,
    \end{cases} \\
    &= (1.0\cdot 10^{-32}\,\mathrm{eV}) \left[\frac{B_0}{10^{-16}\,\mathrm{T}}\right] \left[\frac{|\vec{B}_\mathrm{ext}|}{10^{-10}\,\mathrm{T}}\right]\left[\frac{7.9\,\mathrm{g}\,\mathrm{cm}^{-3}}{\rho}\right] \frac{A}{P}\times\begin{cases}
        \frac{1}{Z}, &\quad \psi = e, \\
        1, &\quad \psi = N,
    \end{cases} 
    \end{split}
\end{equation}
for $R \ll 1$, where we have used $\Delta E_{\psi,B} = 2 \mu_\psi |\vec{B}_\mathrm{ext}|$. For our reference scenario we have chosen $B_0 = 5\cdot 10^{-14}\,\mathrm{T}$, corresponding to the SQUID magnetometer discussed in~\cite{Lamoreaux:2001hb}, and used the density of iron in place of $\rho$. It is clear that the SQUID magnetometer setup is at the very least as sensitive as the torsion balance setup discussed in Section~\ref{sec:torsion}, but can be made more sensitive by decreasing the applied magnetic field. The ideal setup would therefore be to initially apply a strong external magnetic field to align the target spins, and then steadily decrease the applied field to maximise the acquired transverse polarisation. 

One should also notice that $\omega_{\psi,0} = \omega_{\psi,L} \sqrt{1+R^2}$, where $\omega_{\psi,L} = 2\pi f_{\psi,L}$ is the angular Larmor frequency of the system. The overall magnetisation of the system would therefore initially precess about the applied magnetic field with frequency $\omega_{\psi,L}$, increasing to $\sqrt{2}\omega_{\psi,L}$ as the applied field is turned down. This effect could also be interpreted as a field dependent gyromagnetic ratio. Given, however, that $R \ll 1$ for most reasonable scenarios, we do not expect this to have an observable effect on the signal. We show the time dependence of the SQUID magnetometer signal in Figure~\ref{fig:squid}, including the case where the DM wind is not exactly orthogonal to the applied magnetic field. In particular, we show the signal when the fraction of the relative frame velocity along $\vec{B}_\mathrm{ext}$, $\beta_{\Earth,\parallel} = 1/\sqrt{3}$, or equivalently when the relative velocity is split equally along each direction. Importantly, this does not have a drastic effect on magnitude of the signal, and so should not severely impact the sensitivity of this method outside the extreme case where $\beta_{\Earth,\parallel} \to 1$. 

\subsection{Example: a scalar DM model}
To give a rough estimate of the constraints that can be placed on DM using this method, we consider the two component DM model given in~\cite{Boehm:2020wbt}, which features a heavy, leptophilic dark vector mediator $Z'_{\mu}$ and complex scalar $\phi$ with interaction Lagrangian
\begin{equation}
    \mathcal L_{Z'} = g_\phi^2 Z'_\mu Z'^{\mu} |\phi|^2-  i g_\phi \phi^* \overleftrightarrow{\partial_\mu} \phi Z'^{\mu} - Z'_\mu \bar l \gamma^\mu (g_L P_L + g_R P_R)l,
\end{equation}
where $l \in \{e,\mu,\tau\}$, $g_\phi$, $g_L$ and $g_R$ are dimensionless couplings, and $P_{R/L} = (1\pm \gamma^5)/2$ are the right and left chirality projection operators. Focusing on the case with $l = e$, and integrating out the heavy $Z'$ leads to the effective low energy Lagrangian
\begin{equation}\label{eq:smodelEff}
    \mathcal L_{Z'} = -i \frac{g_\phi(g_R + g_L)}{2m_{Z'}^2} \left( \phi^* \overleftrightarrow{\partial_\mu} \phi \right) \bar e \gamma^\mu e -i \frac{g_\phi(g_R - g_L)}{2m_{Z'}^2} \left( \phi^* \overleftrightarrow{\partial_\mu} \phi \right) \bar e \gamma^\mu \gamma^5 e + \dots.
\end{equation}
Of interest to us is the second term, which by comparison with~\eqref{eq:splitphi1} generates an electron energy splitting with magnitude
\begin{equation}
    |\Delta E_e| = \frac{2g_{\phi}|g_R - g_L|}{m_{Z'}^2} \beta_\Earth |n_\phi(\phi) - n_{\phi}(\phi^*)|.
\end{equation}
Next, rewriting $|n_\phi(\phi) - n_{\phi}(\phi^*)| = |\delta_\phi| \rho_\mathrm{DM}/m_\phi$, where $\delta_\phi \in [-1,1]$ parameterises the asymmetry between $\phi$ and $\phi^*$, and considering purely axial couplings, $g_R = -g_L = g_A$, we find
\begin{equation}\label{eq:scalarBoundNFO}
    \begin{split}
        |\Delta E_e| &= \frac{4}{\Lambda_{Z'}^2} \frac{\rho_\mathrm{DM}}{m_\phi} \beta_\Earth |\delta_\phi|. \\
        &= (1.2 \cdot 10^{-34}\,\mathrm{eV}) \left[\frac{\Lambda_{Z'}^{-1}}{356\,\mathrm{TeV}^{-1}}\right]^2 \left[\frac{10\,\mathrm{MeV}}{m_\phi}\right] |\delta_\phi|,
    \end{split}
\end{equation}
for $\beta_\Earth = 7.6\times 10^{-4}$~\cite{Vera:2020}, where we have defined the effective new physics scale $\Lambda_{Z'} = m_{Z'}/\sqrt{g_\phi |g_A|}$ and assumed that $\phi$ makes up the entire local relic density, $\rho_\mathrm{DM} \simeq 0.4\,\mathrm{GeV}\,\mathrm{cm}^{-3}$. If we instead assume production via freeze-out, we can estimate the local DM density of $\phi$ in terms of $\Lambda_{Z'}$ and $m_\phi$, which for $m_{Z'}^2 \gg m_\phi^2 \gg m_e^2$ has a different scaling to~\eqref{eq:scalarBoundNFO}
\begin{equation}\label{eq:scalarBoundFO}
    |\Delta E_e| = (1.2 \cdot 10^{-34}\,\mathrm{eV}) \left[\frac{356\,\mathrm{TeV}^{-1}}{\Lambda_{Z'}^{-1}}\right]^2 \left[\frac{10\,\mathrm{MeV}}{m_\phi}\right]^3 |\delta_\phi|.
\end{equation}
In both cases, the reference value, $\Lambda_{Z'}^{-1} = 356\,\mathrm{TeV}$ corresponds to the approximate value required to reproduce the relic density for $m_\phi = 10\,\mathrm{MeV}$. More generally, we require
\begin{equation}\label{eq:foLimit}
    \Lambda_{Z'}^{-1} \gtrsim (356\,\mathrm{TeV}^{-1})\left[\frac{10\,\mathrm{MeV}}{m_\phi}\right]^{\frac{1}{2}},
\end{equation}
so as not to overclose the universe. It is instructive to recast both~\eqref{eq:scalarBoundNFO} and~\eqref{eq:scalarBoundFO} in terms of the constraints that can be placed on the effective new physics scale using the DSE. Given a sensitivity to energy shifts $|\Delta E_0| \gtrsim 10^{-32}\,\mathrm{eV}$, corresponding to the SQUID magnetometer considered in~\eqref{eq:squidSens}, we find the constraint on the effective new physics scale
\begin{equation}\label{eq:constraintFO}
    \Lambda_{Z'}^{-1} \gtrsim (38.8\,\mathrm{TeV}^{-1})\left[\frac{10\,\mathrm{MeV}}{m_\phi}\right]^\frac{3}{2} \left[\frac{10^{-32}\,\mathrm{eV}}{|\Delta E_0|}\right]^{\frac{1}{2}}\sqrt{|\delta_\phi|},
\end{equation}
assuming the energy splitting from freeze-out production~\eqref{eq:scalarBoundFO}, which for $\mathcal{O}(1)$ values of the asymmetry parameter, \textit{i.e.} supposing that the dark sector matter-antimatter asymmetry follows that of the visible sector, is just one order of magnitude away from being able to probe $\Lambda_{Z'}$ that reproduces the measured relic density at $m_\phi = 10\,\mathrm{MeV}$. If we instead assume that $\phi$ makes up the entirety of dark matter independent of $m_\phi$ and $\Lambda_{Z'}$, corresponding to the energy splitting~\eqref{eq:scalarBoundNFO}, we find the constraint
\begin{equation}\label{eq:constraintNFO}
    \Lambda_{Z'}^{-1} \lesssim (3.3\cdot 10^3\,\mathrm{TeV}^{-1})\left[\frac{m_\phi}{10\,\mathrm{MeV}}\right]^{\frac{1}{2}}\left[\frac{|\Delta E_0|}{10^{-32}\,\mathrm{eV}}\right]^{\frac{1}{2}}\frac{1}{\sqrt{|\delta_\phi|}},
\end{equation}
which is once more roughly an order of magnitude away from the freeze-out band~\eqref{eq:foLimit}. 
\begin{figure}
    \centering
    \includegraphics[width = \textwidth]{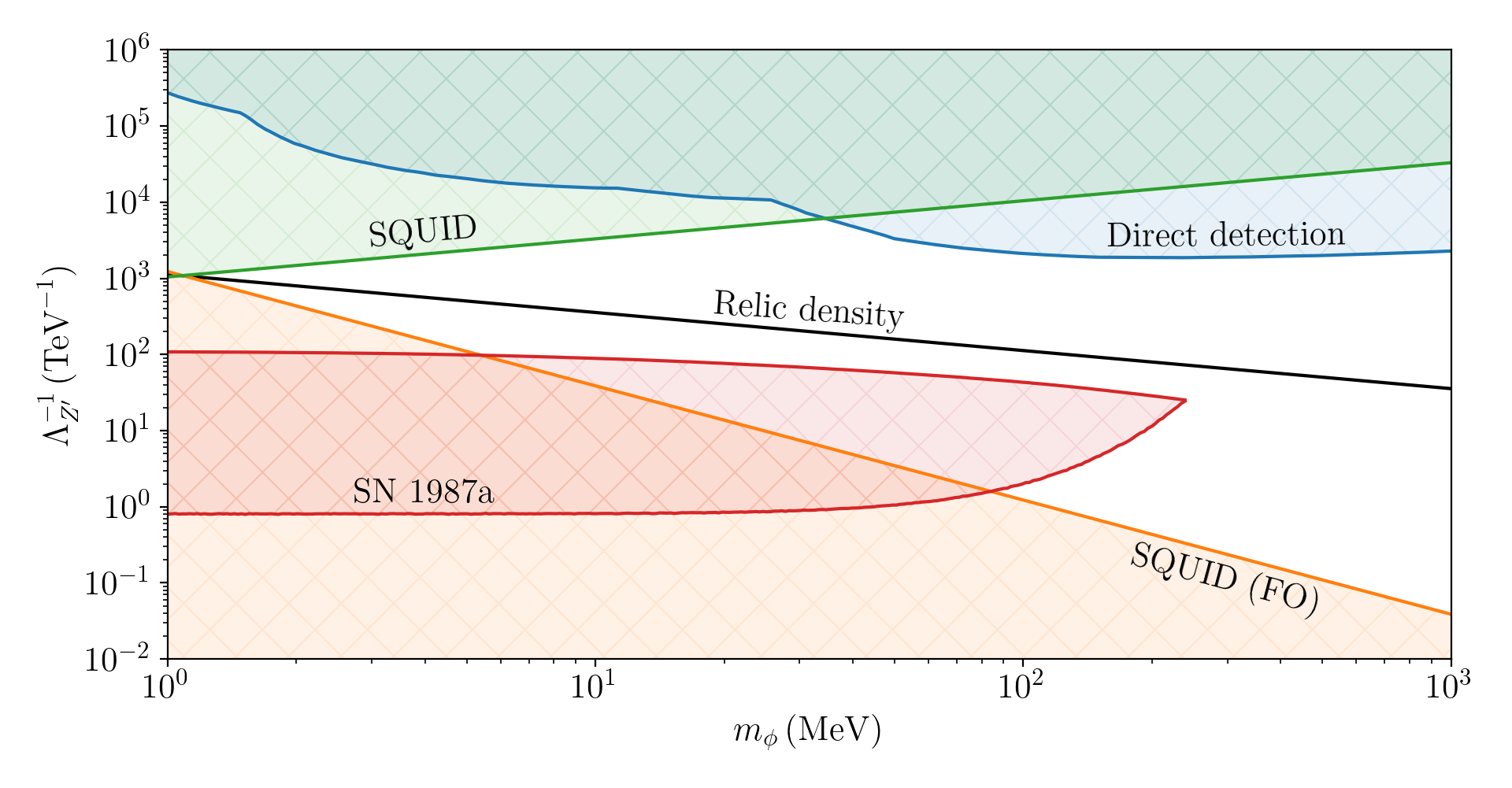}
    \caption{Constraint projections on the effective DM coupling, $\Lambda_{Z'}^{-1} = \sqrt{g_\phi g_A}/m_{Z'}$, from the SQUID magnetometer for the generic (green) and freeze-out (orange) production scenarios, where we assume $\delta_\phi = 1$. We compare these with the constraints from direct detection experiments~\cite{SENSEI:2020dpa,Essig:2017kqs, XENON:2019gfn} (blue), assuming constant DM form factors, and anomalous supernova cooling constraints (red), which we compute following the method of~\cite{Boehm:2020wbt} for the $18\,M_\Sun$ progenitor discussed in~\cite{Fischer:2016cyd}. For comparison, we show the combination of parameters that reproduce the local relic density for a freeze-out scenario with the black curve, corresponding to the saturation of~\eqref{eq:foLimit}.}
    \label{fig:bounds}
\end{figure}
We show the constraints that could be placed on $\Lambda_{Z'}^{-1}$ using a SQUID magnetometer in Figure~\ref{fig:bounds} as a function of $m_\phi$ for both the freeze-out (FO) and unspecified production scenarios, and compare these with the existing constraints from direction detection experiments~\cite{SENSEI:2020dpa,Essig:2017kqs, XENON:2019gfn} and anomalous supernova cooling, computed following the method of~\cite{Boehm:2020wbt}. As expected, this experiment significantly outperforms existing direct detection experiments for $m_\phi \lesssim 30\,\mathrm{MeV}$. Additionally, if freeze-out is assumed, the SQUID magnetometer experiment is instead able to place constraints on the minimum value of $\Lambda_{Z'}^{-1}$, owing to the linear scaling of the DSE with the effective coupling. Importantly, this includes regions that are currently unconstrained by SN 1987a.

Aside, notice that the energy splitting due to $\phi$ backgrounds far exceeds that expected from the C$\nu$B for the parameter ranges considered here, assuming the same asymmetry for both. It is therefore entirely possible that the DSE completely washes out the Stodolsky effect for neutrinos. One could also envisage scenarios in which the opposite is true, and the DSE is overwhelmed by the C$\nu$B, or those in which one acts as a significant background to the other. This should be taken into consideration when using this technique, especially as it is difficult to distinguish between the operators responsible for the DSE. Nevertheless, the observation of either the DSE or Stodolsky effect for neutrinos would be a strong indicator of as-yet-unobserved physics.   
%
%
\section{Conclusions}\label{sec:conclusions}
Despite comprising $\sim 26\%$ of the energy density of the universe, detecting DM is an incredible challenge that has yet to be accomplished. Here we have explored the possibility of constraining DM models using the DSE: tiny energy splittings between the spin states of SM fermions induced by an incoherent DM background. Throughout, we have used an EFT formalism and identified all effective DM operators up to dimension-6, for DM candidates with spin-0 to spin-$\tfrac{3}{2}$, that can give rise to the DSE. Our key finding is that the energy splittings due to the DSE scale linearly with the effective DM coupling, inversely with the DM mass, and are roughly independent of the DM kinematics. Importantly, this differs from traditional DM direct detection experiments, where the sensitivity typically decreases with decreasing DM mass. On the other hand, every operator discussed here requires either a particle-antiparticle or helicity asymmetry in the background to give a non-zero contribution to the DSE. This technique therefore favours chiral models and those with a sizeable chemical potential during production, however we note that either asymmetry may develop post-production through several mechanisms \textit{e.g.} DM reflection at surface of the Earth, scattering on polarised backgrounds. 

In this work, we have identified two methods through which these tiny energy splittings can be observed. The first utilises an extremely sensitive, polarised torsion balance, which experiences a torque due to the energy splittings induced by the DM background. For a conservative setup, this experiment is sensitive to energy splittings of $\Delta E_\psi \simeq 10^{-28}\,\mathrm{eV}$, but could have a sensitivity to splittings as small as $\Delta E_\psi \simeq 10^{-36}\,\mathrm{eV}$ for a more optimistic setup. The second utilises a SQUID magnetometer to detect the time-varying magnetisation of a target due to the DM background, which acts similarly to an external magnetic field on the target. We estimate that this experiment will be sensitive to splittings of $\Delta E_\psi \simeq 10^{-32}\,\mathrm{eV}$. 

Finally, we have explored a scalar DM model, considering both the case where the new scalar constitutes the entire local DM density regardless of the model parameters, and the more realistic scenario where it is produced via freeze-out. In both scenarios, we showed the SQUID magnetometer proposal is able to exclude regions of parameter space that are not already ruled out by direct detection experiments or SN 1987a, provided that there is a sizeable asymmetry in the DM background. For the range of parameters considered, we also demonstrated that the DSE for the scalar DM model far exceeded the Stodolsky effect for neutrinos, provided that the asymmetry in both backgrounds was comparable.  Clearly, the DSE is a powerful tool to constrain DM models in otherwise difficult-to-test regions of parameter space.
%
%
\acknowledgments

We would like to thank Martin Bauer for some very useful comments about the effective operator basis, and Xiao-Dong Ma for highlighting two redundant operators in a previous version of this work. We are also grateful to Yuber F. Perez-Gonzalez, Lucien Heurtier and Animesh Datta for some helpful comments during the preparation of this manuscript. Jack D. Shergold is supported by an STFC studentship under the STFC training grant ST/T506047/1.


\appendix
\section{Lab frame averaging}\label{sec:newAv}
In this appendix we will describe the averaging procedure used to compute the energy shifts in the lab frame. We begin by assuming that the DM is described by an isothermal spherical halo, with galaxy frame velocity distribution
\begin{equation}\label{eq:fGal}
    f(\vec{p}) = \left(\frac{2\pi}{m_\mathrm{DM}^2 \sigma^2}\right)^\frac{3}{2}e^{-\frac{|\vec{p}|^2}{2m_\mathrm{DM}^2\sigma^2}},
\end{equation}
where $\vec{p}$ is the DM momentum in the galactic reference frame, $m_\mathrm{DM}$ is its mass and $\sigma$ is the velocity dispersion. The normalisation factor is found by requiring that $\int \tfrac{d^3p}{(2\pi)^3} f(\vec{p}) = 1$. As a result of the frame transformation, DM particles in the lab frame will not follow~\eqref{eq:fGal} but instead the transformed distribution function $f_\mathrm{lab}$, such that the average of some lab frame quantity $X_\mathrm{lab}$ will be given by
\begin{equation}\label{eq:labav1}
    \left\langle X_\mathrm{lab} \right\rangle = \int \frac{d^3p}{(2\pi)^3}X_\mathrm{lab} f_\mathrm{lab}(\vec{p}) = \frac{1}{(2\pi)^3}\int X_\mathrm{lab} f_\mathrm{lab}(\vec{p}) |\vec{p}|^2 \sin\theta \,d|\vec{p}|\, d\theta \,d\phi.
\end{equation}
To find $f_\mathrm{lab}(\vec{p})$, we first note that since all velocities involved are small, the momentum of the DM particle in the lab frame $\vec{p}_\mathrm{lab}$ can be written in terms of the relative frame velocity $\vec{\beta}_\Earth$ as
\begin{equation}
    \vec{p}_\mathrm{lab} \simeq \vec{p} + m_\mathrm{DM} \vec{\beta}_\Earth = |\vec{p}|{\left(\begin{array}{c}
         \cos\phi \sin\theta  \\
          \sin\phi\sin\theta  \\
          \cos\theta
    \end{array}\right)} + m_\mathrm{DM}\beta_\Earth{\left(\begin{array}{c}
         0  \\
          0  \\
          1
    \end{array}\right)},
\end{equation}
where $\beta_\Earth \equiv |\vec\beta_\Earth|$, and we have chosen $\vec{\beta}_\Earth || z$ for simplicity. This choice makes no difference at the level of averaging, but becomes important when considering experimental setups. We will therefore write our final expressions for averaged quantities in terms of a general orientation of $\vec{\beta}_\Earth$. Next, since $f_\mathrm{lab}(\vec{p}_\mathrm{lab}) = f(\vec{p})$, the lab frame distribution function will satisfy
\begin{equation}
    f_\mathrm{lab}(\vec{p}) = f(\vec{p} - m_\mathrm{DM}\Vec{\beta}_\Earth) = \left(\frac{2\pi}{m_\mathrm{DM}^2 \sigma^2}\right)^\frac{3}{2} e^{-\frac{|\vec{p}|^2 + m_\mathrm{DM}^2 \beta_\Earth}{2m_\mathrm{DM}^2\sigma^2}}e^{\frac{|\vec{p}|\beta_\Earth\cos\theta}{m_\mathrm{DM}\sigma^2}},
\end{equation}
which can be readily plugged into~\eqref{eq:labav1} to compute averaged lab frame quantities. 

In addition to the distribution function, we must also write the lab frame polarisation vectors in terms of DM reference frame quantities. To do so, we rotate the polarisation vectors~\eqref{eq:polVecs} to point along an arbitrary axis, and then use $\vec{p}_\mathrm{lab}$ to rewrite angles in the lab frame in terms of those in the DM frame, yielding
\begin{align}
    \epsilon_+^\mu = (\epsilon_-^\mu)^* &= \frac{1}{\sqrt{2}}{\left(\begin{array}{c}
         0  \\
         \frac{1}{|\vec{p}_\mathrm{lab}|}\cos\phi \left(|\vec{p}| \cos\theta + \beta_\Earth m_\mathrm{DM}\right) - i\sin\phi  \\
         \frac{1}{|\vec{p}_\mathrm{lab}|}\sin\phi \left(|\vec{p}| \cos\theta + \beta_\Earth m_\mathrm{DM}\right) + i\cos\phi \\
         -\frac{|\vec{p}|}{|\vec{p}_\mathrm{lab}|}\sin\theta
    \end{array}\right)}, \\
    \epsilon_L^\mu &= {\left(\begin{array}{c}
         \frac{|\vec{p}_\mathrm{lab}|}{m_\mathrm{DM}}  \\
         \frac{|\vec{p}|}{|\vec{p}_\mathrm{lab}|} \cos\phi \sin\theta  \\
         \frac{|\vec{p}|}{|\vec{p}_\mathrm{lab}|}\sin\phi \sin\theta  \\
          \frac{1}{|\vec{p}_\mathrm{lab}|} \left(|\vec{p}| \cos\theta + \beta_\Earth m_\mathrm{DM}\right)
    \end{array}\right)},
\end{align}
again assuming $\vec\beta_\Earth || z$. 

Relaxing the assumption $\vec\beta_\Earth || z$, we find the averages relevant to the operators considered in this work  
\begin{align}
    \left\langle\frac{1}{E_\mathrm{DM}}(\vec{p}_\mathrm{DM}\cdot \vec{S}_\psi)\right\rangle &= 2\beta_\Earth s_{\psi,\parallel},\label{eq:fav1}\\
    \left\langle\frac{1}{E_\mathrm{DM}}(p_\mathrm{DM}\cdot S_\mathrm{\psi})\right\rangle &= -2\beta_\Earth s_{\psi,\parallel}, \\
    \begin{split}\left\langle\frac{1}{E_\mathrm{DM}}(S_\mathrm{DM}\cdot S_\mathrm{\psi})\right\rangle &= \left[\frac{(1 - 8\beta_r^2)}{8\beta_r^2}\mathrm{Erf}\left(2\beta_r\right)-\frac{1}{2\sqrt{\pi}\beta_r}e^{-4\beta_r^2}\right]\frac{s_{\psi,\parallel}}{m_\mathrm{DM}} \\
    &\simeq -\frac{7}{8}\frac{s_{\psi,\parallel}}{m_\mathrm{DM}} + \mathcal{O}(1-\beta_r), 
    \end{split}\\
    \begin{split}\left\langle\frac{1}{E_\mathrm{DM}}(p_\mathrm{DM}\cdot S_\psi)(S_\mathrm{DM}\cdot p_\psi)\right\rangle &= \Bigg[\frac{(1-16\beta_r^2 - 64\beta_r^4)}{16\beta_r^4}\mathrm{Erf}\left(2\beta_r\right) \\
    & \qquad\qquad\qquad\qquad\qquad- \frac{(1+ 8\beta_r^2)}{4\sqrt{\pi}\beta_r}e^{-4\beta_r^2}\Bigg]\beta_c^2 m_\psi s_{\psi,\parallel}\\
    &\simeq -5 \beta_c^2 m_\psi s_{\psi,\parallel} + \mathcal{O}(1-\beta_r),
    \end{split} \label{eq:fav4}\\
    \begin{split}\left\langle\frac{1}{E_\mathrm{DM}}(p_\mathrm{DM}\cdot p_\psi)(S_\mathrm{DM}\cdot S_\psi)\right\rangle &= \left[\frac{(1 - 8\beta_r^2)}{8\beta_r^2}\mathrm{Erf}\left(2\beta_r\right)-\frac{1}{2\sqrt{\pi}\beta_r}e^{-4\beta_r^2}\right]m_\psi s_{\psi,\parallel} \\
    &\simeq -\frac{7}{8}m_\psi s_{\psi,\parallel} + \mathcal{O}(1-\beta_r), 
    \end{split}\label{eq:fav5} \\
    \left\langle\frac{1}{E_\mathrm{DM}}\varepsilon_{\alpha\beta\mu\nu}p_\mathrm{DM}^\alpha\, p_\psi^\beta S_\mathrm{DM}^\mu S_\psi^\nu\right\rangle &= 0, \\
    \begin{split}\left\langle\frac{1}{E_\mathrm{DM}}\varepsilon_{\alpha\beta\mu\nu}p_\psi^\alpha S_\psi^\beta \epsilon^{*\mu}_\pm\epsilon^\nu_\pm\right\rangle &= \pm i\left[\frac{(1 - 8\beta_r^2)}{8\beta_r^2}\mathrm{Erf}\left(2\beta_r\right)-\frac{1}{2\sqrt{\pi}\beta_r}e^{-4\beta_r^2}\right]\frac{m_\psi}{m_\mathrm{DM}} s_{\psi,\parallel} \\
    &\simeq \mp \frac{7i}{8} \frac{m_\psi}{m_\mathrm{DM}}s_{\psi,\parallel} + \mathcal{O}(1-\beta_r), 
    \end{split}\label{eq:favEpsPol}\\
    \left\langle\frac{1}{E_\mathrm{DM}}\varepsilon_{\alpha\beta\mu\nu}p_\psi^\alpha S_\psi^\beta \epsilon^{*\mu}_L\epsilon^\nu_L\right\rangle &= 0, \\
    \left\langle\frac{1}{E_\mathrm{DM}}(p_\psi \cdot \epsilon_\pm)(S_\psi\cdot \epsilon^*_\pm)\right\rangle &= 0,
    \end{align}
    \begin{align}
    \left\langle\frac{1}{E_\mathrm{DM}}(p_\psi \cdot \epsilon_L)(S_\psi\cdot \epsilon^*_L)\right\rangle &= -\frac{2m_\psi}{m_\mathrm{DM}} \beta_\Earth s_{\psi,\parallel},\label{eq:favLast}\\
    \left\langle\frac{1}{E_\mathrm{DM}}(\vec{p}_X \cdot \vec{\epsilon}_\pm)( \epsilon^*_\pm \cdot S_\psi)\right\rangle &= 0, \\
    \left\langle\frac{1}{E_\mathrm{DM}}(\vec{p}_X \cdot \vec{\epsilon}_L)( \epsilon^*_L \cdot S_\psi)\right\rangle &= -2\beta_\Earth s_{\psi,\parallel}, \\
    \left\langle\frac{1}{E_\mathrm{DM}}\varepsilon_{\alpha\beta i\nu}\epsilon^{*\alpha}_{\pm}\epsilon^\beta_\pm p_X^i S_{\psi}^\nu \right\rangle &= 0, \\
    \left\langle\frac{1}{E_\mathrm{DM}}\varepsilon_{\alpha\beta i\nu}\epsilon^{*\alpha}_L\epsilon^\beta_L p_X^i S_{\psi}^\nu \right\rangle &= 0,
    \end{align}
with $s_{\psi,\parallel} = (\vec{\beta}_\Earth \cdot \vec{s}_\psi)/\beta_\Earth$ and $\beta_r = \beta_\Earth/\beta_c$, where $\beta_c = \sqrt{2}\sigma$ is the circular velocity of the galaxy. The presence of $s_{\psi,\parallel}$ indicates that only the spin state directed along the DM wind experiences an energy shift. We will not include this factor explicitly in the main text. 
%
%
\section{Operator basis}\label{sec:reduction}
Here we outline the identities used to reduce the effective DM operator bases to those appearing in Section~\ref{sec:DMops}. We begin by noting that excluding field indices, the Lorentz structures that can enter into our effective DM operators are the ones already given in~\eqref{eq:gamStrucs}, along with the partial derivative, $\partial_\mu$, and Levi-Civita tensor, $\varepsilon_{\alpha\beta\mu\nu}$. Considering operators up to dimension-6 with a fermionic SM part, we can therefore have at most a single derivative entering. As such, for spin-0 and spin-$\frac 12$ DM particles, the only way that the Levi-Civita tensor can enter a Lagrangian is through operators that contain at least three gamma matrices; which can then be reduced to simpler Lorentz structures via the Chisholm identity and the definition of $\gamma^5$
\begin{align}
    \gamma^{\alpha}\gamma^{\beta}\gamma^{\mu} &= \eta^{\alpha\beta}\gamma^{\mu} + \eta^{\beta\mu}\gamma^{\alpha} - \eta^{\alpha\mu}\gamma^{\beta} - i\varepsilon^{\sigma\alpha\beta\mu}\gamma_{\sigma}\gamma^5, \\
    \gamma^5 &= \frac{i}{4!}\varepsilon_{\alpha\beta\mu\nu} \gamma^\alpha \gamma^\beta \gamma^\mu \gamma^\nu,
\end{align}
where $\eta^{\mu\nu}$ is the metric tensor. These can also be used to derive the identity
\begin{equation}
    \varepsilon_{\alpha\beta\mu\nu} \sigma^{\mu\nu} =  -2i \sigma_{\alpha\beta} \gamma^5,
\end{equation}
which generates the additional Lorentz structure given in~\eqref{eq:sigma5}, allowing us to express operators containing a Levi-Civita tensors in terms of operators containing the more convenient $\sigma^{\mu\nu} \gamma^5$ structure. The trace of the product of fermion spinors with this structure is
\begin{equation}
    \mathrm{Tr} \left[ u_\psi \bar u_\psi \sigma^{\mu\nu} \gamma^5 \right] = 2i\left( p_\psi^\mu S_\psi^\nu - S_\psi^\mu p_\psi^\nu \right).
\end{equation}
Additionally, we can use this to demonstrate that $\mathcal O_{\chi_5} = \mathcal O_{\chi_4}$ via
\begin{equation}
    i(\bar \chi \sigma_{\mu\nu} \gamma^5 \chi)(\bar \psi \sigma^{\mu\nu} \psi) = -\frac 12 \varepsilon_{\mu\nu\alpha\beta} (\bar \chi \sigma^{\alpha\beta} \chi)(\bar \psi \sigma^{\mu\nu} \psi) =  i (\bar \chi \sigma^{\alpha\beta} \chi)(\bar \psi \sigma_{\alpha\beta} \gamma^5 \psi),
\end{equation}
and that $\mathcal O_{\chi_6} = \mathcal O_{\chi_3}$ using
\begin{equation}
    \begin{split}
        (\bar \chi \sigma_{\mu\nu} \gamma^5 \chi)(\bar \psi \sigma^{\mu\nu} \gamma^5 \psi) &= -\frac 14 \varepsilon_{\mu\nu\alpha\beta} \varepsilon^{\mu\nu\delta\gamma} (\bar \chi \sigma^{\alpha\beta} \chi)(\bar \psi \sigma_{\delta\gamma} \psi) \\
        &= \frac 12 \left( \delta_\alpha^\delta  \delta_\beta^\gamma - \delta_\alpha^\gamma \delta_\beta^\delta \right) (\bar \chi \sigma^{\alpha\beta} \chi)(\bar \psi \sigma_{\delta\gamma} \psi) \\
        &= (\bar \chi \sigma^{\alpha\beta} \chi)(\bar \psi \sigma_{\alpha\beta} \psi).
    \end{split}
\end{equation}
Many operators containing derivatives can be reduced using symmetry currents, and through integration by parts. Take, for example, the scalar operator
\begin{equation}
    \left( \partial_\mu |\phi|^2 \right) \bar \psi \gamma^\mu \gamma^5 \psi = \partial_\mu \left[ |\phi|^2 \bar \psi \gamma^\mu \gamma^5 \psi \right] - |\phi|^2 \partial_\mu \left( \bar \psi \gamma^\mu \gamma^5 \psi \right).
\end{equation}
The first term on the right-hand side is a total derivative and so will not contribute to the classical action, whilst the second term contains the derivative of the axial current which can be re-expressed as $2im_\psi |\phi|^2 \bar \psi \gamma^5 \psi$ using the equations of motion for spin-$\frac 12$ fields.  We can therefore perform the operator reduction
\begin{equation}
    \left( \partial_\mu |\phi|^2 \right) \bar{\psi}\gamma^\mu \gamma^5 \psi \longrightarrow |\phi|^2 \bar\psi \gamma^5 \psi,
\end{equation}
which as per~\eqref{eq:traces} does not contribute to the DSE. Note that similar structures containing vector currents vanish from the requirement $\partial_\mu (\bar \psi \gamma^\mu \psi) = 0$. 

Further reductions in the effective operator basis are obtained using equations of motion. In particular, we make use of the spin-1 and spin-$\frac 32$ equations of motion, which lead to the constraints
\begin{equation}
    \partial_\mu X^\mu = 0, \qquad \partial_\mu \Psi^\mu = 0,\qquad \gamma_\mu \Psi^\mu = 0,
\end{equation}
allowing us to eliminate or simplify operators in which spin-1 fields share an index with a derivative. Manipulations using the third identity, $\slashed{\Psi} = 0$, yields the basis given in Table~\ref{tab:rsOperators}.

The operator bases used throughout this work are those in which the equations of motion have been applied maximally. This avoids the need to apply Hamilton's equations to the Hamiltonian when computing the energy shifts, which is a far more involved task than using the Euler-Lagrange equations. For completeness, we also specify the spin-independent members of our operator bases: at spin-0, these are $|\phi|^2 \bar\psi \psi$, $i|\phi|^2 \bar\psi \gamma^5 \psi$, and $i(\phi^\dagger \overleftrightarrow{\partial_\mu}\phi) (\bar\psi\gamma^\mu\psi)$; for spin-1 DM, we have $|X|^2 \bar\psi\psi$, $i|X|^2 \bar\psi\gamma^5\psi$, along with the vector current analogues of each axial-vector operator appearing in Table~\ref{tab:vectorOperators}, hermitianised appropriately with factors of the imaginary unit. Finally, the full spin-$\tfrac{1}{2}$ basis includes products of fermion bilinears not given in Table~\ref{tab:fermionOperators}, whilst the complete basis for RS fermions is given in~\cite{Ding:2012sm}. The same spin-0 basis, along with a similar spin-1 basis can also be found in~\cite{He:2022ljo}. 
\section{Levi-Civita identity}\label{sec:lcIdentity}
Here we derive an identity that can be used to evaluate contractions of four-vectors and a Levi-Civita tensor of the form
\begin{equation}
    \varepsilon_{\alpha\beta\mu\nu} A^\alpha B^\beta C^\mu D^\nu,
\end{equation}
in a more practical manner, where $A,B,C$ and $D$ are some unspecified four-vectors. First we note that this contraction will always contain at least three spatial components, and so will carry a global factor of $(-1)^3$ regardless of the four-vectors considered. Next, recalling that we can write a contraction of the Levi-Civita symbol and a series of vectors as a matrix determinant, we have
\begin{equation}
    \varepsilon_{\alpha\beta\mu\nu} A^\alpha B^\beta C^\mu D^\nu = (-1)^3 \left|{\begin{array}{cccc}
        A_0 & A_1 & A_2 & A_3  \\
        B_0 & B_1 & B_2 & B_3  \\
        C_0 & C_1 & C_2 & C_3  \\
        D_0 & D_1 & D_2 & D_3   
    \end{array}}\right|,
\end{equation}
which can be neatly re-expressed in terms of scalar triple products as
\begin{equation}\label{eq:triple}
    \varepsilon_{\alpha\beta\mu\nu} A^\alpha B^\beta C^\mu D^\nu = - A_0 [\Vec{B},\Vec{C},\Vec{D}] + B_0 [\Vec{A},\Vec{C},\Vec{D}] - C_0 [\Vec{A},\Vec{B},\Vec{D}] + D_0 [\Vec{A},\Vec{B},\Vec{C}],
\end{equation}
where $[\Vec{A},\Vec{B},\Vec{C}] = \Vec{A}\cdot (\Vec{B}\times \Vec{C})$ is the scalar triple product which is unchanged by cyclic permutations, and antisymmetric under the interchange of any two elements. One could also choose to take the determinant in other ways which may better suit the experimental setup.  

To demonstrate the use of this identity, we consider a contraction that may occur between a dark fermion $\chi$ and a SM fermion $\psi$, which have four-momenta and spin, $p$ and $S$, respectively
\begin{equation}\label{eq:epsContractSM}
    \varepsilon_{\alpha\beta\mu\nu} p^\alpha_\chi p^\beta_\psi  S^\mu_\chi S^\nu_\psi.
\end{equation}
If the SM fermion is at rest in the lab frame, we can use~\eqref{eq:triple} to reduce~\eqref{eq:epsContractSM} to
\begin{equation}
    \varepsilon_{\alpha\beta\mu\nu} p^\alpha_\chi p^\beta_\psi  S^\mu_\chi S^\nu_\psi = m_\psi [\Vec{p}_\chi, \Vec{S}_\chi, \Vec{S}_\psi].
\end{equation}
In the most general case, this can then be evaluated by explicitly plugging in values for the relevant spins and momenta. However, if we instead consider the DM to be in a helicity eigenstate, we will have $\vec{S}_\chi \parallel \vec{p}_\chi$, from which it follows that
\begin{equation}
    \varepsilon_{\alpha\beta\mu\nu} p^\alpha_\chi p^\beta_\psi  S^\mu_\chi S^\nu_\psi = 0,
\end{equation}
by making use of the antisymmetric property of the scalar triple product.
%
%
\section{Fermion spin precession}\label{sec:squidDeriv}
Here we derive the spin precession of an SM fermion in a combined magnetic and DM background field that gives rise to the transverse magnetisation~\eqref{eq:magnetisation}. To do so, we need to set up the differential equation that governs the evolution of the SM fermion spin. There will be two components to this: the precession due to the DM background, and the precession due to an external magnetic field. Both of these are due to the same effect, a non-diagonal Hamiltonian resulting from the energy splittings due to background fields. We begin with the time-dependent Schr\"odinger equation, which for our system takes the form 
\begin{equation}\label{eq:tdse}
    i\frac{\partial}{\partial t} \psi(x,t) = (H_\mathrm{kin}(x) + V_\mathrm{DM} + V_B)\psi(x,t),
\end{equation}
where $\psi(x,t)$ is the fermion wavefunction, $H_\mathrm{kin}(x)$ is its kinetic Hamiltonian, which is spin and time-independent, whilst $V_\mathrm{DM}$ and $V_B$ are the potentials due to the DM background and applied magnetic field, respectively, which are spin-dependent and we will treat as constant in time here\footnote{In truth, at least one of these must be time-dependent. If we fix our coordinate system in the lab frame, then due to the relative motion of the Earth to the DM reference frame, the direction of the background wind will change in time. However, this can alternatively be accounted for by weighting the collected data by the projection of the relative velocity onto the magnetic field direction. See supplementary material S10.1 of~\cite{Garcon:2019inh} for details of the weighting, and~\cite{Bandyopadhyay:2010zj} for a full parametrisation of the relevant coordinate systems.}. This motivates the factorisation
\begin{equation}
    \psi(x,t) = X(x) T(t),
\end{equation}
where $X(x)$ is a scalar, containing the spatial components of the wavefunction, and $T(t)$ is an eigenspinor of the form
\begin{equation}
    T(t) = {\left(\begin{array}{c}
         T_+(t)  \\
         T_-(t) 
    \end{array}\right)}, \quad |T(t)|^2 = 1.
\end{equation}
This factorisation makes~\eqref{eq:tdse} separable, but it is easier to note that 
\begin{equation}
    H_\mathrm{kin}(x) X(x) = E_\mathrm{kin} X(x),
\end{equation}
such that we can absorb $E_\mathrm{kin}$ as a time-independent, spin-diagonal contribution to the potential. The overall factor of $X(x)$ can then be factored out, allowing us to write
\begin{equation}\label{eq:separated}
    \left(i\frac{\partial }{\partial t} - H\right) T(t) = 0,
\end{equation}
where $H$ is the total Hamiltonian, including the spin-diagonal contribution from $H_\mathrm{kin}$. If the magnetic field is defined such that it points along $z$, then the $z$ oriented spin state will experience an energy shift. Additionally, the up and down spin states should experience a shift of opposite sign. The potential due to the magnetic field should therefore be proportional to the spin operator along $z$, that is
\begin{equation}
    V_B = \frac{\Delta E_{\psi,B}}{2} S_z = \frac{\Delta E_{\psi,B}}{2}{\left(\begin{array}{cc}
         1 & 0 \\
         0 & -1 
    \end{array}\right)},
\end{equation}
where $\Delta E_B$ is the energy shift due to the magnetic field. We see that this has the desired properties, as if we act on an $S_z$ eigenstate with eigenvalue\footnote{We adopt the convention $S_i = \sigma_i$, with $i \in \{x,y,z\}$ and $\sigma$ denoting a Pauli matrix.} $s_z = \pm 1$, we get the eigenvalue $s_z \Delta E_B/2$. Next, we seek to do the same for the potential due to the DM, which should be directed along the DM wind. Explicitly,
\begin{equation}
    V_\mathrm{DM} = \frac{\Delta E_\psi}{2}\left(\beta_{\Earth,x} S_x + \beta_{\Earth,y} S_y + \beta_{\Earth,z} S_z\right) = \frac{\Delta E_\psi}{2}{\left(\begin{array}{cc}
         \beta_{\Earth,z} & \beta_{\Earth,x}- i\beta_{\Earth,y} \\
         \beta_{\Earth,x} + i\beta_{\Earth,y} & -\beta_{\Earth,z} 
    \end{array}\right)},
\end{equation}
where $\beta_{\Earth,i} = (\vec{\beta}_\Earth \cdot \vec{e}_i) / \beta_\Earth \in [-1,1]$ is the fraction of the relative frame velocity along the direction $i$. We should also include a diagonal term due to the spin-independent effects from the DM, however this can simply be absorbed into $E_\mathrm{kin}$. The total Hamiltonian is then
\begin{equation}
    H = \frac{1}{2}{\left(\begin{array}{cc}
         2E_\mathrm{kin} + \Delta E_{\psi,B}+ \Delta E_\psi \beta_{\Earth,z} & \Delta E_\psi(\beta_{\Earth,x} - i\beta_{\Earth,y}) \\
         \Delta E_\psi(\beta_{\Earth,x} + i\beta_{\Earth,y})  & 2E_\mathrm{kin} - \Delta E_{\psi,B} - \Delta E_\psi\beta_{\Earth,z}
    \end{array}\right)},
\end{equation}
such that the solution to~\eqref{eq:separated} is given by
\begin{equation}\label{eq:xipm}
    T_\pm(t) = \left[s_\pm\cos\left(\frac{\omega_\psi}{2}t\right) - \frac{(\beta_{\Earth,x} \mp i\beta_{\Earth,y})R s_\mp \pm (1 \pm \beta_{\Earth,z} R) s_\pm}{\sqrt{1+2\beta_{\Earth,z}R + R^2}}i\sin\left(\frac{\omega_\psi}{2}t\right)\right] e^{-i E_\mathrm{kin}t},
\end{equation}
where $\omega_\psi = \Delta E_{\psi,B}\sqrt{1 + 2\beta_{\Earth,z}R + R^2}$ is the (angular) precession frequency of the system, proportional to the Larmor frequency, $s_\pm = T_\pm(0)$ are the initial values of the SM fermion eigenspinor, and $R = \Delta E_{\psi}/\Delta E_{\psi,B} \in [-1,1]$ is the ratio of the energy shifts due to each of the background potentials. We further note that $s_+$ is always real, whilst $s_-$ may be complex.

To compute the spin precession using $T_\pm$, we note that the time derivative of some operator $\mathcal{O}$ is given by Heisenberg's equation of motion
\begin{equation}
    \frac{d\mathcal{O}}{dt} = i[H,\mathcal{O}],
\end{equation}
such that the time derivative of each of the expectation values, $s_i$, is
\begin{equation}
    \frac{ds_i}{dt} = T(t)^\dagger \left(\frac{dS_i}{dt} \right) T(t).
\end{equation}
Plugging in~\eqref{eq:xipm}, we find
\begin{align}
    \frac{ds_x}{dt} &= \left(\frac{\beta_{\Earth,x} (1 + \beta_{\Earth,z} R)}{\sqrt{1+2\beta_{\Earth,z}R + R^2}} \sin(\omega_\psi t) + \beta_{\Earth,y}\cos(\omega_\psi t)\right)\Delta E_\psi,\label{eq:spinEvox} \\
    \frac{ds_y}{dt} &= \left(\frac{\beta_{\Earth,y} (1 + \beta_{\Earth,z} R)}{\sqrt{1+2\beta_{\Earth,z}R + R^2}} \sin(\omega_\psi t) - \beta_{\Earth,x}\cos(\omega_\psi t)\right)\Delta E_\psi,\label{eq:spinEvoy} \\
    \frac{ds_z}{dt} &= \left(-\frac{(1 - \beta_{\Earth,z}^2)R}{\sqrt{1+2\beta_{\Earth,z}R + R^2}} \sin(\omega_\psi t)\right)\Delta E_\psi,\label{eq:spinEvoz}
\end{align}
where we have related $s_\pm$ to the initial values of $s_x, s_y$ and $s_z$, using
\begin{align}
    s_{x,0} &= T(0)^\dagger S_x \,T(0) = 2 s_+ \,\mathrm{Re}\left(s_-\right) = 0, \\
    s_{y,0} &= T(0)^\dagger S_y \,T(0) = 2 s_+ \,\mathrm{Im}\left(s_-\right) = 0, \\
    s_{z,0} &= T(0)^\dagger S_z \,T(0) = |s_+|^2 - |s_-|^2 = 1,
\end{align}
and assumed that the spins are initially aligned with the external magnetic field. Notice that all three of~\eqref{eq:spinEvox},~\eqref{eq:spinEvoy} and~\eqref{eq:spinEvoz}, especially those along $x$ and $y$, are proportional to the energy splitting due to the background DM field, and so will vanish in its absence. These equations are readily solved to find the expectation values of the SM fermion spin as a function of time
\begin{align}
    s_x(t) &= \frac{2 R}{\sqrt{1+2\beta_{\Earth,z}R + R^2}}\left[\frac{\beta_{\Earth,x}(1 + \beta_{\Earth,z}R)}{\sqrt{1+2\beta_{\Earth,z}R + R^2}}\sin^2\left(\frac{\omega_\psi}{2}t\right) + \frac{\beta_{\Earth,y}}{2}\sin\left(\omega_\psi t\right)\right], \\
    s_y(t) &= \frac{2 R}{\sqrt{1+2\beta_{\Earth,z}R + R^2}}\left[\frac{\beta_{\Earth,y}(1 + \beta_{\Earth,z}R)}{\sqrt{1+2\beta_{\Earth,z}R + R^2}}\sin^2\left(\frac{\omega_\psi}{2}t\right) - \frac{\beta_{\Earth,x}}{2}\sin\left(\omega_\psi t\right)\right], \\
    s_z(t) &= 1-\frac{2R^2(1 - \beta_{\Earth,z}^2)}{1+2\beta_{\Earth,z}R + R^2}\sin^2\left(\frac{\omega_\psi}{2}t\right),
\end{align}
such that the magnitude of the spin along the transverse direction evolves according to
\begin{equation}\label{eq:sperp}
    \begin{split}
        |s_\perp(t)| &= \sqrt{s_x(t)^2 + s_y(t)^2} \\
        &= \frac{2 |R\sin\left(\frac{\omega_\psi}{2}t\right)|}{1+2\beta_{\Earth,z}R + R^2} \sqrt{1 - \beta_{\Earth,z}^2}\\
        &\qquad\qquad\qquad\times\sqrt{1+2\beta_{\Earth,z}R + R^2\left[\cos^2\left(\frac{\omega_\psi}{2}t\right) +\beta_{\Earth,z}^2 \sin^2\left(\frac{\omega_\psi}{2}t\right)\right]},
    \end{split}
\end{equation}
which vanishes identically when $|\beta_{\Earth,z}| = 1$, or equivalently when the DM wind is colinear with the magnetic field. Consequently, the expression equivalent to~\eqref{eq:sperp} for a general magnetic field orientation is found by making the replacement $\beta_{\Earth,z} \to \beta_{\Earth,\parallel}$, where $\beta_{\Earth,\parallel}$ is the fraction of the relative frame velocity along the external magnetic field direction. The corresponding transverse magnetisation is simply $|M_\perp(t)| = n_\psi \mu_\psi |s_\perp(t)|$, where $n_\psi$ is the number density of SM fermions in the target, and $\mu_\psi$ is their magnetic moment. Given that $\omega_{\psi,0} = \Delta E_{\psi,B}\sqrt{1+R^2}$, this reduces to~\eqref{eq:magnetisation} when $\beta_{\Earth,\parallel} = 0$. For completeness, we note that~\eqref{eq:sperp} has a maximum of
\begin{equation}
    |s_\perp(t_\mathrm{max})| = \frac{2|R|\sqrt{1-\beta_{\Earth,\parallel}^2}\sqrt{1+\beta_{\Earth,\parallel}R + \beta_{\Earth,\parallel}^2 R^2}}{1+\beta_{\Earth,\parallel}R +  R^2}, \quad t_\mathrm{max} = \frac{(2k+1)\pi}{\omega_\psi},
\end{equation}
where $k \in \{0,1,2,\dots\}$. This recovers~\eqref{eq:mmax} for $\beta_{\Earth,\parallel} = 0$.
%
%
\bibliographystyle{JHEP}
\bibliography{bibliography}



\end{document}